\documentclass[preprint,preprintnumbers,amsmath,amssymb]{revtex4}

\usepackage{graphicx}
\usepackage{dcolumn}
\usepackage{bm}

\newcommand\nn         {\nonumber}

\newcommand\as         {\ensuremath{\alpha_{\mathrm{s}}}}

\newcommand\Oe[1]      {\ensuremath{\mathrm O(\ep^{#1})}}

\newcommand{\cV}       {{\cal V}}

\def\ep{\epsilon}
\def\beq{\begin{equation}}
\def\eeq{\end{equation}}
\def\beqn{\begin{eqnarray}}
\def\eeqn{\end{eqnarray}}

\def\bom#1{{\mbox{\boldmath $#1$}}}
\def\to{\rightarrow}

\newcommand{\la}{\langle}
\newcommand{\ra}{\rangle}

\def\nn{\nonumber}

\def\ID{1 \kern -.45 em 1}

\def\Li{\mathop{\mathrm{Li}}\nolimits}

\def\mathswitchr#1{\relax\ifmmode{\mathrm{#1}}\else$\mathrm{#1}$\fi}

\newcommand{\tautb}{\tau_{tb}}
\newcommand{\tautg}{\tau_{tg}}
\newcommand{\taugb}{\tau_{gb}}
\newcommand{\ttb}{t_{tb}}
\newcommand{\ttg}{t_{tg}}

\newcommand{\xn}{{N}}
\newcommand{\mt}{m_{t}}
\newcommand{\msq}{m^2_{t}}
\newcommand{\qtb}{q_{tb}}
\newcommand{\qtg}{q_{tg}}

\newcommand{\qsqhat}{\hat{Q}}
\newcommand{\qsq}{Q}

\newcommand{\ltb}{l_{tb}}
\newcommand{\ltg}{l_{tg}}
\newcommand{\ptbv}{p_T^{\, {\bar b}~\rm veto}}

\def\draftdate{\relax}
\def\mda{\relax}
\def\mua{\relax}
\def\mla{\relax}
\def\draft{
\def\thtystars{******************************}
\def\sixtystars{\thtystars\thtystars}
\typeout{}
\typeout{\sixtystars**}
\typeout{* Draft mode!
         For final version remove \protect\draft\space in source file *}
\typeout{\sixtystars**}
\typeout{}
\def\draftdate{\today}
\def\mua{\marginpar[\boldmath\hfil$\uparrow$]%
                   {\boldmath$\uparrow$\hfil}%
                    \typeout{marginpar: $\uparrow$}\ignorespaces}
\def\mda{\marginpar[\boldmath\hfil$\downarrow$]%
                   {\boldmath$\downarrow$\hfil}%
                    \typeout{marginpar: $\downarrow$}\ignorespaces}
\def\mla{\marginpar[\boldmath\hfil$\rightarrow$]%
                   {\boldmath$\leftarrow $\hfil}%
                    \typeout{marginpar: $\leftrightarrow$}\ignorespaces}
\overfullrule 5pt
\oddsidemargin -15mm
\marginparwidth 29mm
}

\def\stars{\strut\leaders\hbox{*}\hfill\strut}
\def\starline{\hfil\strut\hfil\hbox to \textwidth {\stars}\hfil}

\def\bentarrow{\:\raisebox{1.3ex}{\rlap{$\vert$}}\!\rightarrow}

\def\bothdk#1#2#3#4#5{
\begin{array}{r c l}
#1 & \rightarrow & #2#3 \\
 & & \:\raisebox{1.3ex}{\rlap{$\vert$}}\raisebox{-0.5ex}{$\vert$}%
\phantom{#2}\!\bentarrow #4 \\
 & & \bentarrow #5
\end{array}
}
\begin{document}
\preprint{CERN-PH-TH/2005-110}
\preprint{DSF-18/2005}
\preprint{hep-ph/0506289}
\title{Next-to-leading order corrections to $Wt$ production and decay}
\author{John Campbell}
\email{John.Campbell@cern.ch}
\affiliation{
Department of Physics, 
 TH Divison, CERN, CH-1211 Geneva 23, Switzerland}
\author{Francesco Tramontano}
\email{Francesco.Tramontano@na.infn.it}
\affiliation{Dipartimento di Scienze Fisiche, Universit\`a di Napoli,\\
``Federico II'' e INFN sezione di Napoli,\\
Complesso di Monte S. Angelo, Napoli, Italy}

\date{\today}

\begin{abstract}
We present the results of a next-to-leading order
calculation of $Wt$ production, including the decays of both
the top quark and the $W$ boson. The effects of radiation in the
decay of the top quark are also included.
The separation of diagrams which appear in the real corrections, into
singly- and doubly-resonant contributions, is performed using a $b$-jet
veto which is motivated by the use of the bottom quark distribution
function. We find that, for a choice of scale which is suitable for
this approach, the QCD corrections are very mild and only change the cross
section by up to $10$\% at the LHC, depending on the severity of the $b$-jet
veto. When further cuts are applied, applicable for a Higgs boson search
in the $H \to W W^\star$ channel, we find that the radiative effects
greatly decrease the number of background events expected from this
process. In addition, the shapes of relevant distributions can be
significantly changed at next-to-leading order.
\end{abstract}

\maketitle

\section{Introduction}

At the LHC, the top quark will be produced copiously in many channels.
As well as the $t{\bar t}$ pair production channel, the top quark
may be produced singly in association with other particles. The rates
for these processes will be sufficient to both study the properties of
the top quark in detail and to provide a significant source of
background events for other analyses~\cite{Altarelli:2000ye}.

In this paper we will discuss the calculation of the next-to-leading
order (NLO) corrections to the production of a single top quark in association
with a $W$ boson. This calculation has been included in the general
purpose NLO program
MCFM~\cite{Campbell:1999ah,Campbell:2000bg,Campbell:2002tg}.
The lowest order process which we consider is,
\beq
\label{eq:wtdecay}
\bothdk{b+g}{W^-+}{t}{\nu+e^++b}{e^-+{\bar \nu}}
\eeq
so that the leptonic decays of the $W^-$ and of the top quark are included.
We note that both at the Tevatron and the LHC, the rate for the charge-conjugate
process involving a $W^+$ and a ${\bar t}$ quark is identical~\footnote{
At the Tevatron, this is due to the fact that the machine is a proton
anti-proton collider. At the LHC, the equality is because
the perturbatively-derived bottom quark distribution functions that we
use are the same for ${\bar b}$ and $b$ quarks.}.
This process has previously been considered extensively at leading
order~\cite{Belyaev:1998dn,Tait:1999cf, Belyaev:2000me}. However, it is
only at next-to-leading order that we obtain accurate predictions of
event rates which are sensitive to the structure of jets in the final
state. Such NLO calculations have so far been available only for the case
where the decays of the heavy quark and $W$ boson are not
included~\cite{Giele:1995kr,Zhu:2002uj}. 

We have extended these predictions to include not only the full spin correlations in the
decays of the $W$ boson and the top quark, but also to include the effects of gluon
radiation in the top quark decay. This is achieved using the same method that has previously
been applied to other single top production channels~\cite{Campbell:2004ch} and which is briefly
described, together with other details of the calculation, in Section~\ref{sec:calculation}.

At next-to-leading order some of the contributions representing the emission of
an additional parton require special attention. One finds that the corrections
involving two gluons in the initial state contain diagrams that would normally
be assigned to the lowest order calculation of the doubly resonant $t{\bar t}$
production process. We discuss our treatment of this complication
in Section~\ref{sec:separation}. 

The results of our calculation are presented in
Section~\ref{sec:results}. We discuss the NLO corrections
at the Tevatron and the LHC, comparing our findings with those obtained
previously in the literature. We also provide updated predictions
obtained using the latest experimental inputs and examine the effect 
of including gluon radiation in the decay of the top quark.

Section~\ref{sec:pheno} contains a study of our results in the context
of the search for an intermediate mass Higgs boson at the LHC. In this
channel the Higgs decays via $WW^\star$, with the final state
containing leptons and missing transverse momentum. Since the Higgs
mass cannot then be reconstructed, theoretical input as accurate as
possible is imperative. To that end, in this
section we apply realistic acceptance and search cuts to all the
final state particles, then compare the
effect of the NLO corrections with the more inclusive results already
presented.

\section{Calculational details}
\label{sec:calculation}

To evaluate the matrix elements for the production and decay of a
$W$-top system, we follow the same strategy as in a previous
calculation of single top production~\cite{Campbell:2004ch}, which is
based on two approximations. The first is that the top quark is
produced and decays exactly on-shell, motivated by the fact that
diagrams without an on shell top quark are suppressed by a factor of
$\Gamma_t/m_t \approx 1\%$. This enables a division of the process into
production and decay stages, with the further approximation that the
interference between radiation in the two stages is neglected. On
general grounds the contribution of these interference terms can be
shown to be of the order of $\alpha_s\,\Gamma_t/m_t$ due to the large
difference between the characteristic time scales of the production
($m_t^{-1}$) and decay ($\Gamma_t^{-1}$) stages. More technical details
and further references can be found in Ref.~\cite{Campbell:2004ch}.

\begin{figure}[!ht]
\begin{center}
\includegraphics[angle=0,width=0.69\columnwidth]{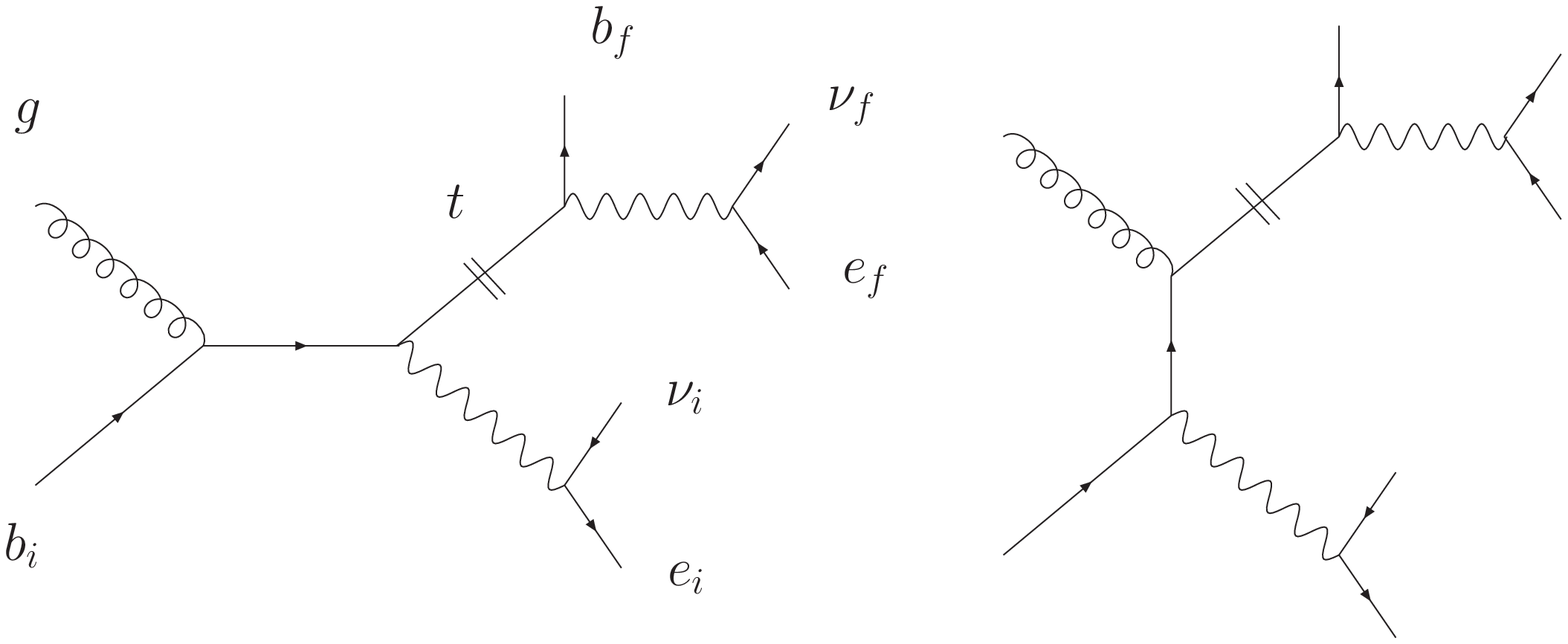}
\caption{The two tree level diagrams for the $Wt$ process. 
\label{fig:TL}}
\end{center}
\end{figure}

\begin{figure}[!ht]
\begin{center}
\includegraphics[angle=0,width=0.55\columnwidth]{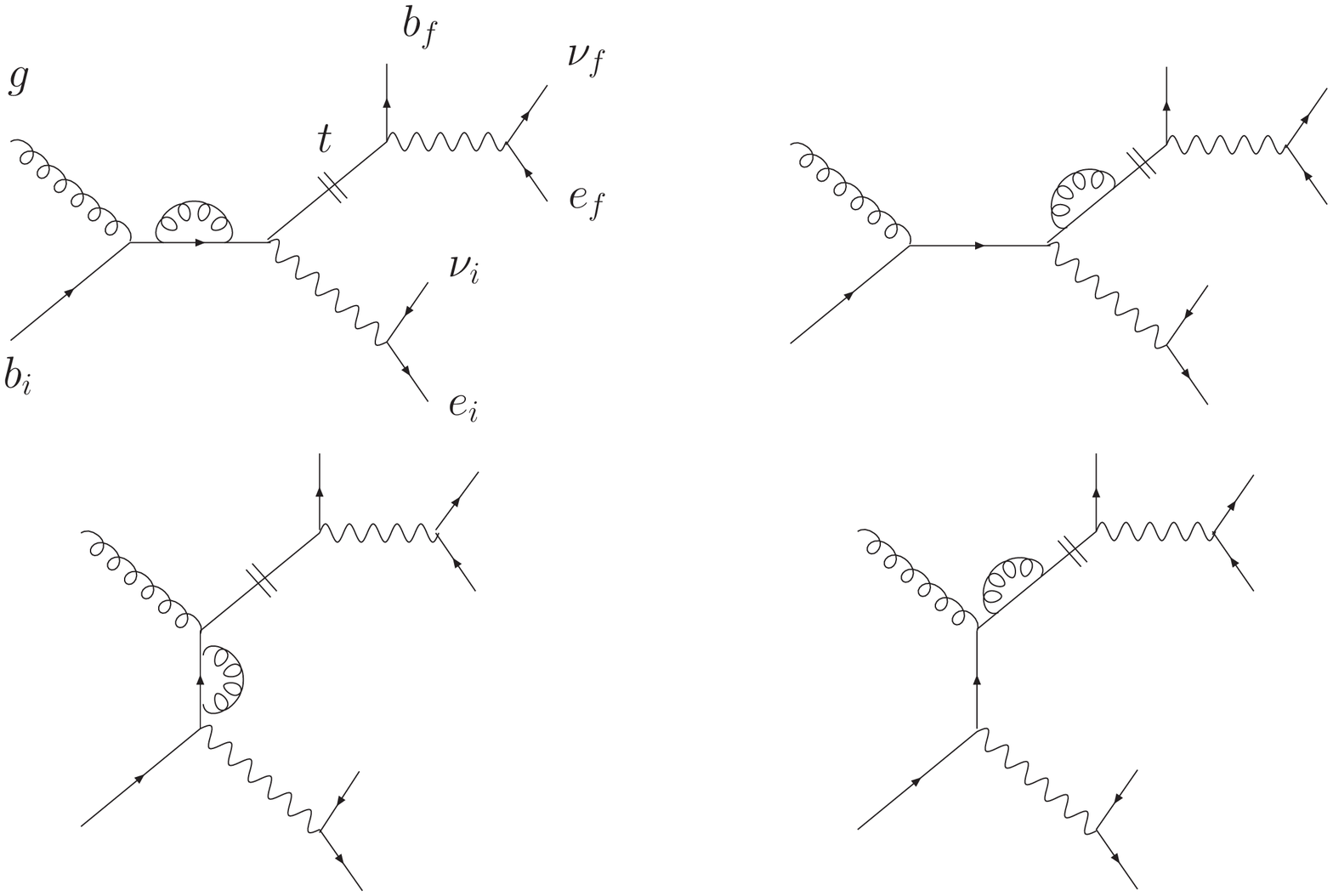}
\includegraphics[angle=0,width=0.75\columnwidth]{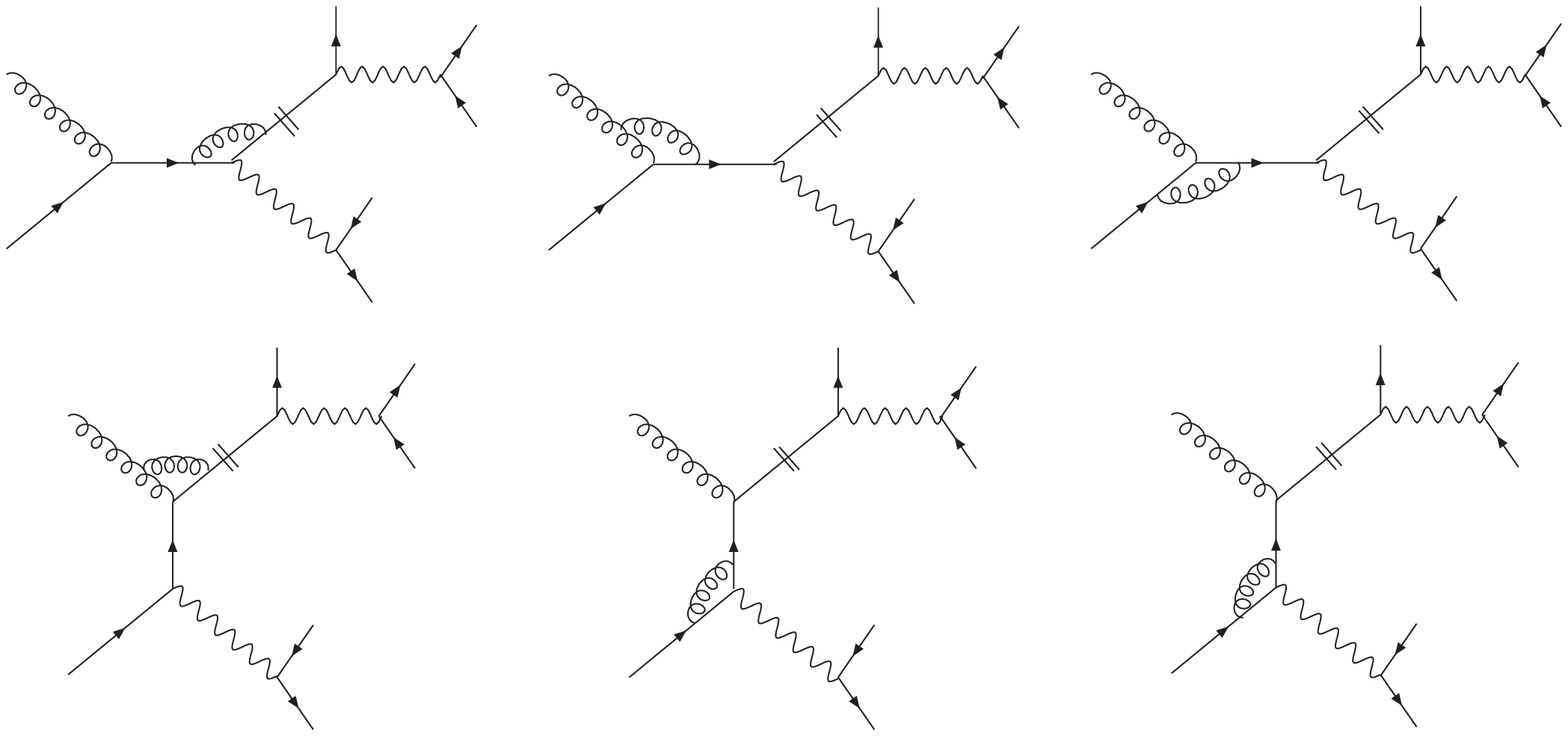}
\includegraphics[angle=0,width=0.75\columnwidth]{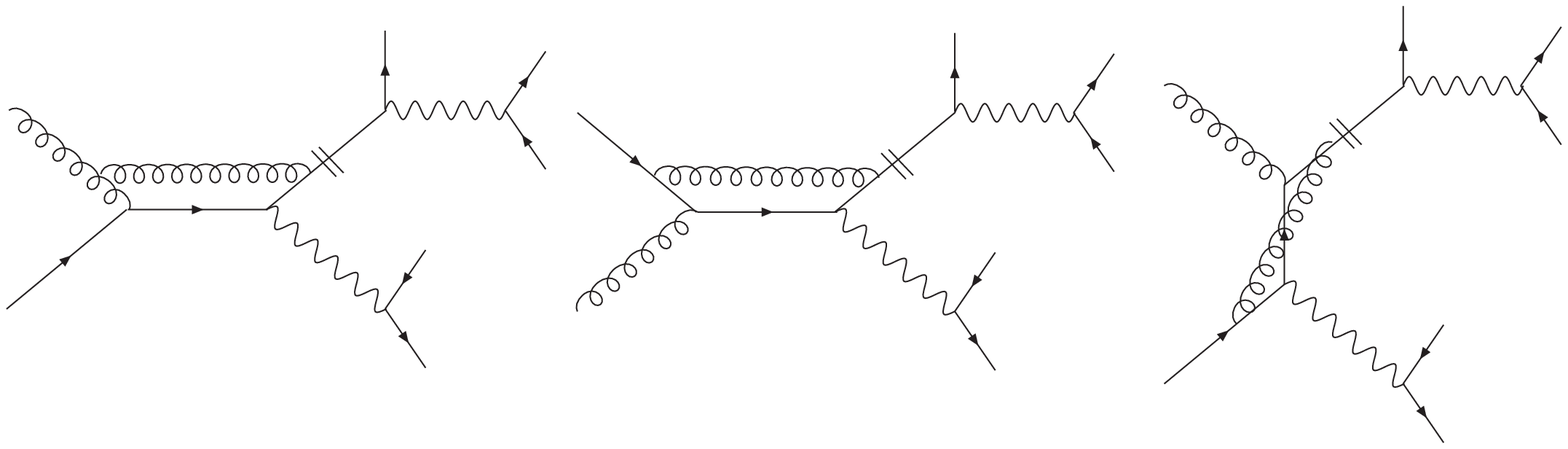}
\caption{One loop diagrams representing virtual corrections in the production stage.
\label{fig:virtp}}
\end{center}
\end{figure}

\begin{figure}[!ht]
\begin{center}
\includegraphics[angle=0,width=0.50\columnwidth]{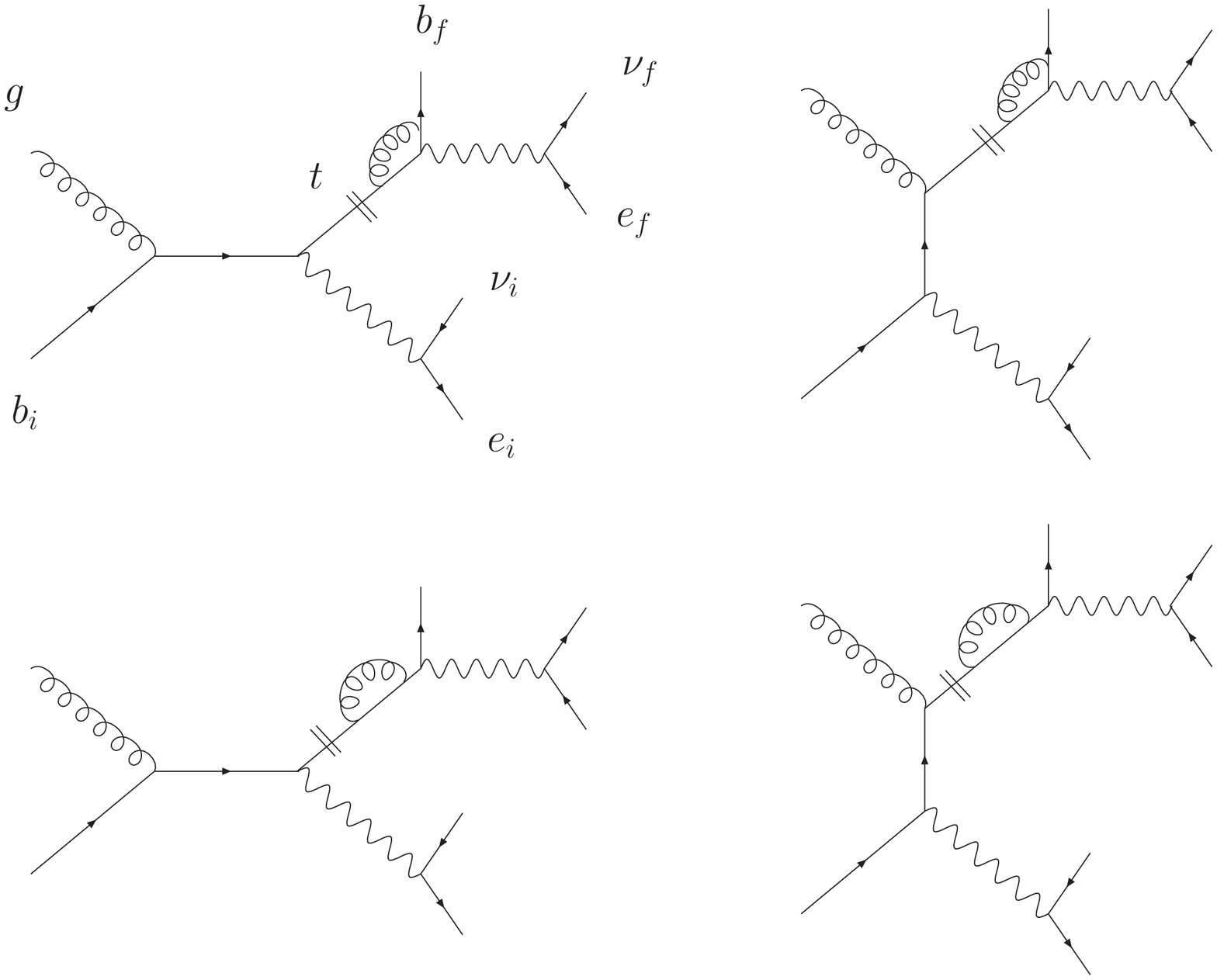}
\caption{One loop diagrams contributing to the calculation of virtual radiation in the decay stage.
\label{fig:virtd}}
\end{center}
\end{figure}

\begin{figure}[!ht]
\begin{center}
\includegraphics[angle=0,width=1.00\columnwidth]{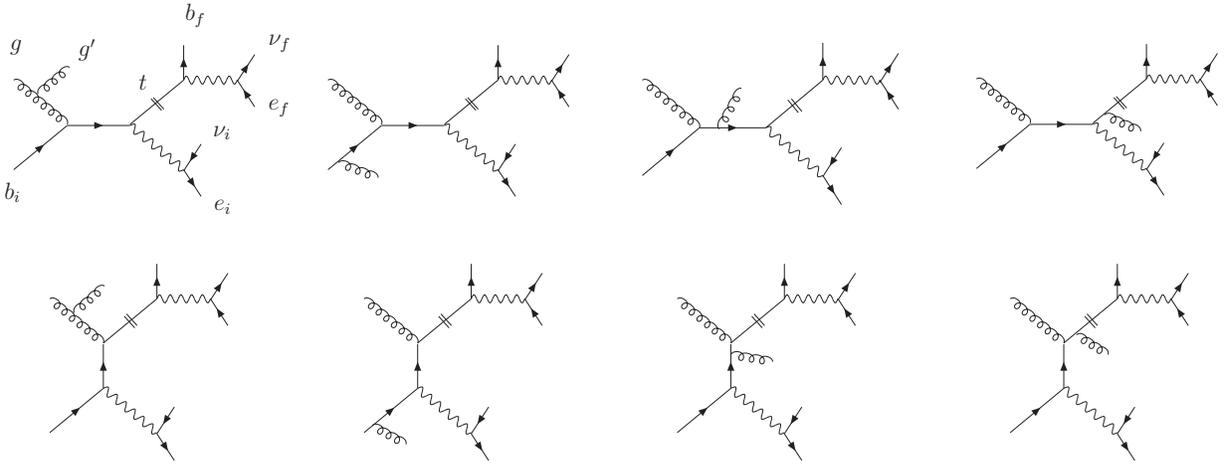}
\caption{Diagrams representing the emission of real radiation in the production stage. In the
calculation, the additional gluon must also be crossed into the initial state and diagrams containing
two quark lines (not shown) are also included.
\label{fig:realp}}
\end{center}
\end{figure}

\begin{figure}[!ht]
\begin{center}
\includegraphics[angle=0,width=0.50\columnwidth]{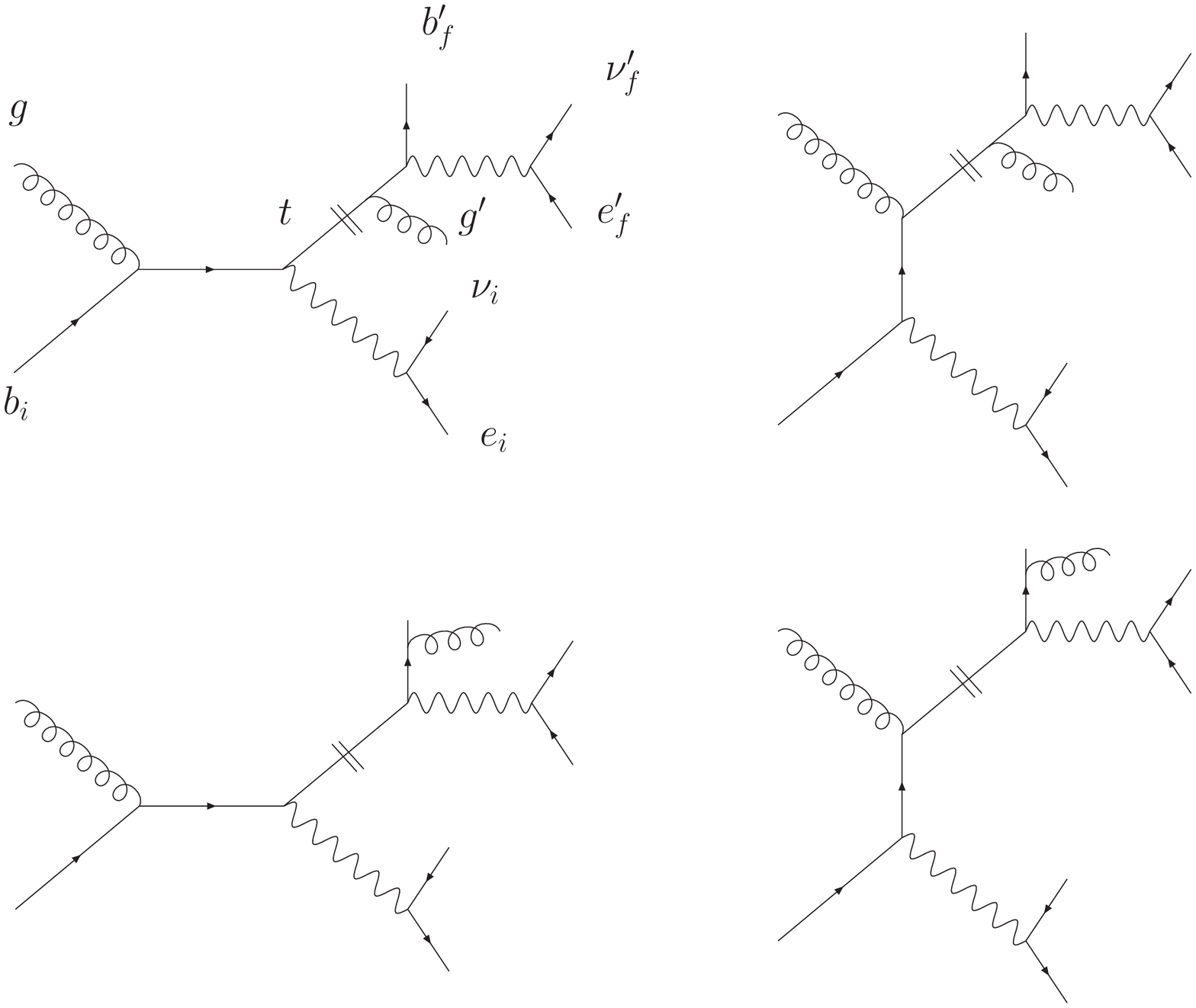}
\caption{Diagrams contributing to real radiation in the decay of the top quark.
\label{fig:reald}}
\end{center}
\end{figure}

The tree level amplitude is represented in Fig.~\ref{fig:TL}, where the on-shell
top propagator is denoted by two short lines.
Due to the weak vertices the only two non-vanishing helicity amplitudes correspond to
the two polarizations of the gluon, with all massless fermions left-handed.
In our expressions for the amplitudes we take all momenta outgoing, restoring incoming
momenta in our implementation by performing the proper analytical continuation of the massless
spinor products.
We write the amplitudes in terms of the momenta of the decay products,
with labels as follows,
\beq
\bothdk{b_i+g}{W^-+}{t}{\nu_f+e_f+b_f}{e_i+\nu_i}
\eeq
and also use $t$ to represent the momentum of the top quark, so that
$t=\nu_f+e_f+b_f$. Both bottom quarks are treated as
massless particles in our approach.
The two tree-level amplitudes are then given by,
\beqn
A_-&=& \frac{f}{\left[\,g\,b_i\,\right]} \,
  \left( \,m_t^2\,\la\,g\,e_i\,\ra\,\left[\,b_i\,e_f\,\right]
- \la\,g\,|\,t\,|\,e_f\,\ra\,\la\,e_i\,\nu_i\,\ra\,\left[\,b_i\,\nu_i\,\right] \right), \nn \\
A_+&=& \frac{f}{\la\,g\,b_i\,\ra} \, 
  \left( \la\,b_i\,|\,t\,|\,e_f\,\ra \, \la\,e_i\,|\,t\,|\,g\,\ra
 - m_t^2\,\la\,b_i\,e_i\,\ra\,\left[\,g\,e_f\,\right]
 + \,2\,g \cdot t\,\la\,e_i\,|\,t\,|\,e_f\,\ra\,
   \frac{\left[\,g\,\nu_i\,\right]}{\left[\,b_i\,\nu_i\,\right]} \right),
\eeqn
where the overall factor $f$ is,
\beq
f = \frac{g_s\,g_w^4\, T^a\; \la\,\nu_f\,b_f\,\ra\,\left[\,b_i\,\nu_i\,\right]}
{\sqrt{2}\,g \cdot t\; W^+_{prop}\,W^-_{prop}\,t_{prop}}.
\eeq
In these formulae $P_{prop}= P^2-M_P^2+iM_P\Gamma_P$,
$T^a$ represents one of the eight $SU(3)$ color generators and
the spinor products with massless four-momenta $p_i$ and $p_j$
are defined as usual:
\beqn
\la \, p_i \,  p_j \, \ra &=&  \la \, p_i - |\, p_j \, + \ra , \\
\left[ \, p_i \, p_j \, \right] &=& \la \, p_i + |\, p_j \, - \ra , \\
\la \, p_i\,|\,p_k\,|\,p_j\,\ra &=& \la\, p_i - |\, \rlap{/}p_k \,|\, p_j - \ra .
\eeqn
As explained above, the virtual corrections to this process can
be divided into production stage corrections (represented
in Fig.~\ref{fig:virtp}) and decay stage ones (shown in Fig.~\ref{fig:virtd}).
The same division applies to the real corrections, which are depicted in Figs.~\ref{fig:realp}
and~\ref{fig:reald} respectively. By producing the top quark strictly on shell we are assured
that the diagrams in Figs.~\ref{fig:realp} and~\ref{fig:reald} are separately gauge
invariant.

The cancellation of the soft and collinear divergences between real and
virtual contributions has been implemented through the subtraction method~\cite{Ellis:1980wv}.
In particular, for the contribution from real radiation in the production stage we have adopted an
extension of the dipole subtraction scheme~\cite{Catani:1996jh,Catani:1997vz} which handles the
case of massive quarks in the final state~\cite{Catani:2002hc}. We have used a
generalization of this method, where one can use a tuneable parameter in order to have
better control over the size of the subtraction. Further details and formulae are
contained in Ref.~\cite{Campbell:2004ch} and Appendix~\ref{appendix:real}.
For the case of real radiation in the decay we have used a further extension of this method,
as in Ref.~\cite{Campbell:2004ch}, which ensures that the top quark and the $W$ boson remain
on-shell when the subtraction is performed.

The virtual corrections in the decay stage amount
to the study of the decay of an on-shell top quark. These amplitudes
have been known for a long time~\cite{Gottschalk:1980rv,Schmidt:1995mr} and we do
not report them here.
To evaluate the contribution of the virtual radiation in the production
stage (Fig.~\ref{fig:virtp}) we start from the amplitudes where the top quark
is produced on-shell without decaying and calculate amplitudes for
the two polarization states of the top quark. This is achieved by writing the
spinors in terms of an auxiliary massless four-vector $t_0$ in the following way:
\beqn
u(t)_\uparrow &=&
\frac{(\rlap{/}t+m_t)}{\left[\,t_0\,g\,\right]}\,|\,g,-1\,\ra \nn , \qquad
\bar{u}(t)_\uparrow =
\la\,g,-1|\,\frac{(\rlap{/}t+m_t)}{\la\,g\,t_0\,\ra} \nn , \\
u(t)_\downarrow &=& \frac{(\rlap{/}t+m_t)}{\la\,t_0\,g\,\ra}\,|\,g,+1\,\ra , \qquad
\bar{u}(t)_\downarrow =
\la\,g,+1|\,\frac{(\rlap{/}t+m_t)}{\left[\,g\,t_0\,\right]} .
\eeqn
The vector $t_0$ is constructed by forming a linear combination of $t$ and $g$,
\beq
t_0^\mu=t^\mu-\frac{m_t^2}{2\,t \cdot g}\,g^\mu .
\eeq
The full result, where the decay of the top quark is included, can then
be obtained by combining these amplitudes with the ones for the
decay $t \to Wb$, calculated in the same way. This is possible since
the intermediate top quark propagator is recovered via the identity,
\beq
u(t)_\uparrow \bar{u}(t)_\uparrow +
u(t)_\downarrow \bar{u}(t)_\downarrow = \rlap{/}t + m_t .
\eeq
Performing the calculation in this factorized way has a number of
advantages. First, useful consistency  checks can be performed using
the amplitudes without the top quark decay. Second, by replacing the
top quark mass appropriately, these amplitudes can be used to study
other processes where this decay is not relevant.

The diagrams of Fig.~\ref{fig:virtp} have been written and algebraically
manipulated using FORM, after we have used the background field gauge to contain the number
of terms generated by the three gluon vertex.
We deal with infrared and ultraviolet singularities by using dimensional
regularization in the four dimensional helicity scheme and 
use the method of Ref.~\cite{Pittau:1997mv} to write the amplitudes in terms
of traces. The appearance of $\gamma_5$ in the weak vertex is then handled by cyclically rotating
these traces so that $\gamma_5$ appears at the beginning of each trace, before
performing the contraction of Lorentz indices.
Using this prescription we have checked that the Ward identity for the weak current
is satisfied exactly and no additional counter terms are required.

Finally, we are left with box vector integrals and triangle rank 2 tensor integrals.
Using Passarino-Veltman $n$-dimensional decomposition, we obtain a result in terms
of scalar integrals. Due to the nature of our approach, other spurious divergences
are still present at this stage. Individual terms in the result appear to be divergent as
factors in the denominator approach zero. However, in this limit, a combination of such
apparently-singular terms is finite. 
By collecting all such terms over a common denominator, one can identify new functions
that are well-behaved in these limits. These are combinations of rational
and logarithmic functions, as in Ref.~\cite{Campbell:1996zw}. Following this procedure, we
are able to refine our first result considerably.
To show the level of simplification that we have reached,
one of the amplitudes is reported in Appendix~\ref{appendix:virt}.
The others are slightly larger in size and we do not reproduce them here. 
They will be available, together with the rest of our calculation, as part of the
next release of the Monte Carlo program MCFM.

\section{Separation of $Wt$ and $t{\bar t}$ diagrams}
\label{sec:separation}

\begin{figure}
\begin{center}
\includegraphics[angle=0,width=0.45\columnwidth]{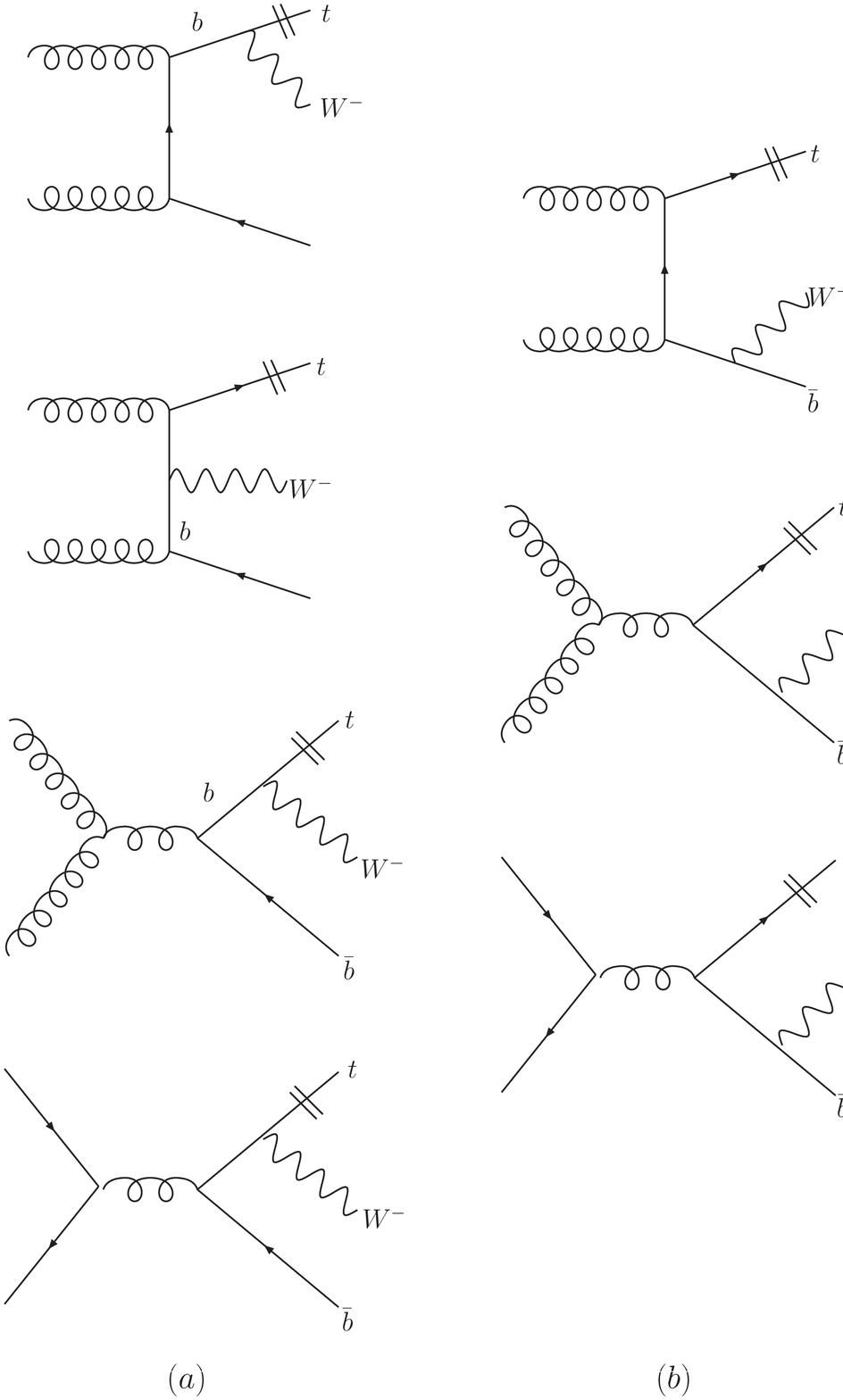}
\caption{Real corrections to $W^-t$ production which involve an
additional ${\bar b}$ quark. The double bars indicate the on-shell
top quark which subsequently decays into $W^+b$. Diagrams obtained by
interchanging two gluons are not shown. The 3 representative diagrams
in the right panel (b) contain a resonant ${\bar t}$ propagator, while
those on the left (a) do not.
\label{fig:wtbdiags}}
\end{center}
\end{figure}

When calculating the real radiation corrections, all appropriate
crossings of the diagrams shown in Fig.~\ref{fig:realp} should be
included. Some of the crossings, in which the additional parton is a
${\bar b}$ quark in the final state, are particularly problematic.
These diagrams are shown in Fig.~\ref{fig:wtbdiags} and involve
gluon-gluon and same flavour quark-antiquark initial states, which are
important at the LHC and Tevatron respectively.
All these diagrams produce a final state consisting of a $W^-$, an
on-shell top quark and a ${\bar b}$ quark. However, the diagrams  in
panel (b) contain a resonant ${\bar t}$ propagator and represent the
production of a $t{\bar t}$ pair with the subsequent decay of the
${\bar t}$ into the $W^-$ and ${\bar b}$ quark. As such, the
contribution from these diagrams when integrated over the total
available phase space can be much larger than the lowest order $Wt$
cross section (an order of magnitude at the LHC). In order to
disentangle these two processes, two methods have been outlined in the
literature.

The first involves making a cut on the invariant mass of the $W^-{\bar
b}$ system to prevent the ${\bar t}$ propagator from becoming
resonant~\cite{Belyaev:1998dn}. The second method instead subtracts the
contribution from the resonant diagrams so that no on-shell piece
remains~\cite{Tait:1999cf}. A comparison of these two
approaches~\cite{Belyaev:2000me} shows that the methods yield the same
total cross section when a mass window of $15 \Gamma_t \approx 25$~GeV
either side of the top mass is chosen. However, these methods do not
lend themselves to a Monte Carlo implementation where one wishes to
study distributions of final state particles as well as total cross
sections. Therefore we shall adopt neither of these prescriptions but
instead follow a procedure motivated by our use of the bottom quark
PDF.

In the $b$-PDF approach, the $b$ quark distribution function 
is derived perturbatively from a collinear $g \to b{\bar b}$ splitting
that occurs in the initial state. It implicitly
includes all splittings up to a $p_T$ of the ${\bar b}$-quark
equal to the factorization scale, $\mu_F$. This means that the
contribution from the corresponding $gg \to Wtb$ diagrams
(contained in panel (a) of Fig.~\ref{fig:wtbdiags})
has already been included in the lowest order
calculation. Therefore the net contribution from these diagrams,
including appropriate counter-terms and integrating over all ${\bar b}$
quark transverse momenta up to $\mu_F$, should be approximately zero.
For a suitable choice of $\mu_F$ we have checked that this is indeed
the case. The choice of $\mu_F$ is made such that the collinear
approximation used in deriving the $b$-PDF is accurate, which for this
process implies that
$\mu_F \lesssim \, (m_W+m_t)/4 \, \approx 65$~GeV~\footnote{
This can be seen by examining a study of the similar process, charged Higgs
production in association with a top quark~\cite{Boos:2003yi}.
We have also reproduced this result for the process at hand.}.

There is also a contribution from the diagrams in panel (b) of Fig.~\ref{fig:wtbdiags}
when the $p_T$ of the ${\bar b}$ quark is small, $p_T^{\bar b} < \mu_F$.
However, these diagrams simply represent the doubly-resonant
$t{\bar t}$ production process which is currently known
up to NLO~\footnote{
A NLO calculation including the decays of the top quarks is not
currently available. However one could, for instance, use a lowest
order calculation including the decays and normalize to the inclusive
$t{\bar t}$ NLO rate, which is known~\cite{Nason:1987xz}.}.
Therefore it is preferable to separate this contribution from the
`genuine' NLO corrections to the $Wt$ process. As we shall show later,
although the contribution from the $t{\bar t}$ diagrams in this region
of phase space is rather small in relation to the total cross section,
it is still competitive with the $Wt$ result. This suppression means
that the interference effects between the two sets of diagrams is very
small when using a $b$-jet veto, in contrast to the case when $p_T^b$
is unconstrained~\cite{Belyaev:2000me}.

When a ${\bar b}$ quark is observed with a $p_T$ above $\mu_F$ then our description
of the final state is a lowest order one. The contribution from the doubly-resonant diagrams
dominates and, as above, a better prediction would be obtained by using the $t{\bar t}$
process.
Alternatively, one could use a calculation including all the diagrams for 
 $gg \to tWb$, including the $t \to bW \to b\ell\nu$ decay (retaining the $b$ quark mass)
and also finite width effects~\cite{Kauer:2001sp}. However, currently this study
would be limited to leading order in $\alpha_s$ only.

To summarize, we shall perform our calculation of the $Wt$ process by
applying a veto on the $p_T$ of the additional $b$ quark that appears
at next-to-leading order. This aids the separation of this process from
doubly-resonant $t{\bar t}$ production. When applying this veto, one
should choose the factorization scale equal to (or at least of the same
order as) the maximum $p_T$ of the $b$ quark that is allowed, $\ptbv$.
This choice respects the approximations that were originally used to
define the $b$ quark PDF. For $\mu_F \ne \ptbv$ and for less inclusive
quantities, the contribution from the $gg \to Wtb$ diagrams is
calculated by simply omitting the doubly-resonant diagrams~\footnote{
As we have already pointed out, interference effects
between singly- and doubly-resonant diagrams are small. Although this
procedure is not strictly gauge invariant, it is no
more serious an error than that incurred when introducing a
Breit-Wigner width for resonant propagators.}.
The result for this piece remains at the level of a few percent of
the lowest order cross section.

\section{Results}
\label{sec:results}

Before discussing the effect of including radiation in the decay of the
top quark, we will first consider just the $Wt$ total cross section in
order to discuss some features of our approach and to compare our
results with those available in the literature.

\subsection{Comparison with no top quark decay}

The NLO corrections to the total $Wt$ cross section, where no decay
of the top quark is included, were previously
presented in Ref.~\cite{Zhu:2002uj}. For the sake of comparison, in
this section we will adopt the parameters therein as closely as
possible. In particular, we choose $m_t=175$~GeV and the CTEQ5 set of
parton distribution functions~\footnote{We use CTEQ5L1 for the lowest
order calculation and CTEQ5M1 at NLO, the versions that include the
improved evolution code.}. The other electroweak parameters that
enter our calculation are chosen to be,
\beq
M_W=80.419~{\mathrm GeV}, \qquad g_W^2=0.4267.
\eeq

We perform our comparison at the LHC and examine the dependence of our
results on the common renormalization and factorization scale $\mu$. As we
have already argued, the $b$-PDF approach is most well-motivated when
choosing a value of $\mu$ less than about $65$~GeV.  However, such a value is
much smaller than the more typical choice, $\mu=m_t+m_W$ which is the central
value chosen in Ref.~\cite{Zhu:2002uj}. Therefore, for the sake of
illustration, we choose to study the scale dependence over a large range
from $\mu=25$ GeV up to $\mu=m_t+m_W=255$~GeV. As we have discussed above, we
limit the $p_T$ of the $b$ quark that appears at next-to-leading order to
have a maximum value $\ptbv$, which we choose here to be $50$~GeV.

\begin{figure}
\begin{center}
\includegraphics[angle=90,width=0.8\columnwidth]{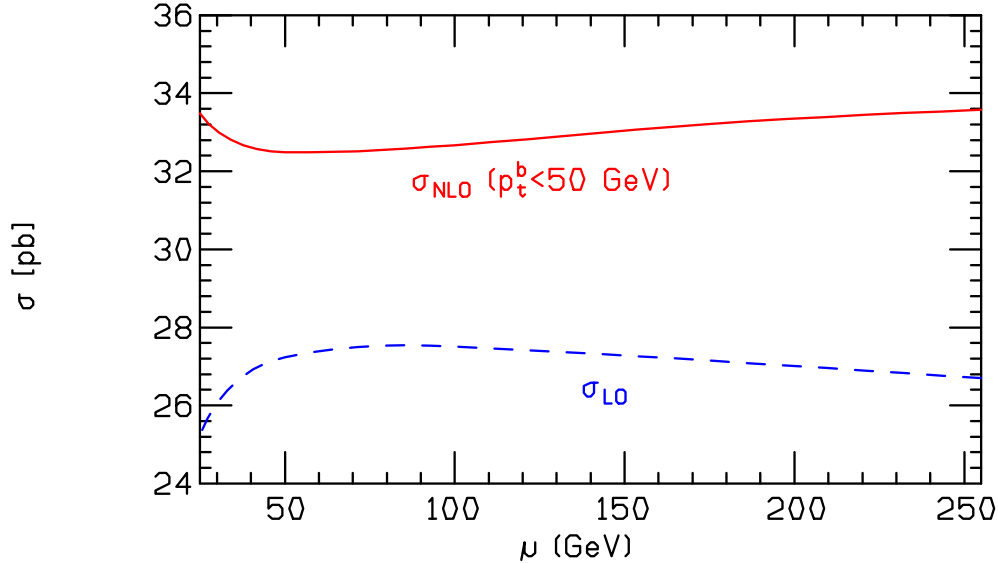}
\caption{Scale dependence of the cross sections for $W^-t$ production at the LHC
for $m_t=175$ GeV. The branching ratios for the decay of the top quark
and the $W$'s are not included. Cross sections are evaluated with
CTEQ5L1 ($\as(M_Z)=0.127$) and CTEQ5M1 ($\as(M_Z)=0.118$)
PDFs~\cite{Lai:1999wy}. We choose a single common renormalization and factorization
scale, $\mu$. The lowest order cross section is the dashed curve, whilst the NLO one
-- calculated with $\ptbv=50$~GeV -- is solid.
\label{fig:mudep_veto50}}
\end{center}
\end{figure}

Our results at LO and NLO are shown in Fig.~\ref{fig:mudep_veto50}.
We see that the dependence of the lowest order curve on a common scale choice
is already remarkably small. At next-to-leading order we see that this is
improved still further, with the cross section varying by about $3\%$ over
the range of scales shown in the figure.

Comparing with Zhu~\cite{Zhu:2002uj}, we find a number of differences. Even
at lowest order our result for $\mu=m_t+m_W$ is higher and furthermore, the
dependence of the result on this scale appears less mild (c.f. Fig.~2 of
Ref.~\cite{Zhu:2002uj}). However, we have checked the lowest order results of
our program against those obtained with MadEvent~\cite{Maltoni:2002qb} and found
good agreement. At next-to-leading order we also find a slightly different
result, lower and with a stronger dependence on the scale. In this case, we
expect some discrepancy due to our different method of handling the $gg \to
Wtb$ contribution. We note that the combination of the ${\bar b}$-jet veto and our
preferred choice of a much smaller scale, leads to a next-to-leading order
cross section that is about $15\%$ smaller than that found by Zhu. In
addition, the $K$-factor, defined as the ratio of NLO and LO cross sections,
is much smaller and in the range $1.2$--$1.3$ depending on the scale choice.

\begin{figure}
\begin{center}
\includegraphics[angle=90,width=0.8\columnwidth]{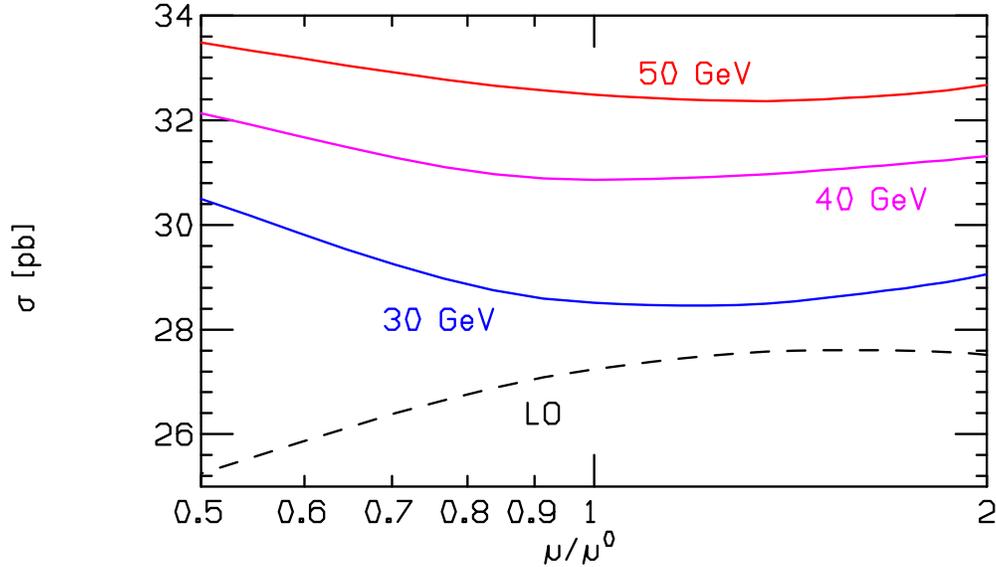}
\caption{Scale dependence of the cross sections for $W^-t$ production at the LHC,
for three different choices of $\ptbv = \mu_0$. From bottom to top,
the solid curves represent values of $30$, $40$ and $50$~GeV. The lowest order dashed
curve (calculated with $\mu_0=50$~GeV) is also shown for reference.
\label{fig:mudep_ptdep}}
\end{center}
\end{figure}

Finally, we consider the dependence of our result on the choice for $\ptbv$. In
Fig.~\ref{fig:mudep_ptdep} we show the scale dependence for three different choices of the
veto threshold. 
We have concentrated on the region of smaller scales and varied the common
scales by a factor of two about the central value $\mu_0=\ptbv$. One sees
that, within this window, the scale dependence of the next-to-leading order
calculation is  again somewhat smaller than that found at LO and improves as
the veto threshold is raised.  Compared to a threshold of $50$~GeV, the
cross section decreases by about $15\%$ when lowering the veto to $30$~GeV.
This substantially redues the effect of the next-to-leading order corrections
on the cross section, leaving a $K$-factor close to unity for our central
scale choice. 

\subsection{Updated results and radiation in the decay}

First, we repeat the calculation of the total $Wt$ cross section, but using
the most recent determination of the top quark mass~\cite{Azzi:2004rc}, which
yields $m_t=178$~GeV. We also use recent PDF sets from the MRST and CTEQ
groups.

Our predictions for the Tevatron and the LHC are shown in
Table~\ref{tab:total}, where we have used $\ptbv=50$~GeV and factorization
and renormalization scales also equal to this value.
It is clear from the quoted cross sections that this process is of
little phenomenological relevance at the Fermilab collider, although
we include the result here for completeness.
\begin{table}[tb]
\caption{\small LO and NLO cross sections for single top-quark
production in association with a $W^-$ at the Tevatron and LHC, for $m_t=178$ GeV. 
The branching ratios for the decays of the top quark and the $W^-$ are not included. 
Cross sections are evaluated with CTEQ6L1 ($\as(M_Z)=0.130$) and CTEQ6M
($\as(M_Z)=0.118$) PDFs~\cite{Pumplin:2002vw} and also with the MRST2002 NLO PDF
set ($\as(M_Z)=0.1197$)~\cite{Martin:2002aw}. 
The renormalization and factorization
scales are set to the ${\bar b}$ jet veto threshold of $50$~GeV.  The errors represent Monte Carlo statistics only.
\label{tab:total}}
\begin{center}
\begin{ruledtabular}
\begin{tabular}{llll} 
Collider, $\sqrt{s}$   & PDF      & $\sigma_{LO}$ [pb] & $\sigma_{NLO}$ [pb] \\
\hline
$p\bar p$, 1.96 TeV    & CTEQ6    & 0.04796            & 0.06458   $\pm$ 0.0001 \\
$p\bar p$, 1.96 TeV    & MRST2002 & 0.08083            & 0.07414   $\pm$ 0.0001 \\
\hline
$pp$,        14 TeV    & CTEQ6    & 29.41 	       & 32.10	$\pm$ 0.03 \\
$pp$,        14 TeV    & MRST2002 & 31.08 	       & 34.49	$\pm$ 0.03 \\
\end{tabular}
\end{ruledtabular}
\end{center}
\end{table}
The NLO corrections at the Tevatron increase the cross section by a
factor of $1.35$ when using the CTEQ PDF set, but decrease it by a little
under $10$\% for the 
MRST parametrization. The NLO results still differ by about 
$15$\%, reflecting the considerable uncertainty in the gluon
distribution at large $x$. At the LHC the effect of the NLO corrections
is much smaller, increasing the cross section by about $10\%$ in both cases,
and the predictions from the two PDF sets show much better agreement. 

We now turn to a fuller description of the final state, where all
the leptonic decays of the top quark and the $W$ bosons are included,
as in Eq.~(\ref{eq:wtdecay}). 
The decays are included using Breit-Wigner propagators
with widths,
\beq
\Gamma_W =2.06~{\mathrm GeV}, \qquad
\Gamma_t^{\mathrm LO}=1.651~{\mathrm GeV},
\eeq
and we now investigate the effects of the inclusion of gluon radiation in the
decay of the top quark. Although the inclusion of this radiation should not
change the total cross section, a difference is expected when working at a
fixed order of perturbation theory. Our results are summarized in
Table~\ref{tab:decay}.
\begin{table*}[tb]
\caption{\small Comparison of LO and NLO cross sections for $W^-t$
production at the Tevatron and LHC, with leptonic decays of both the $W^-$ and the
top quark. The NLO calculation is performed both
without including QCD effects in the decay ($\sigma B_{t \to b \nu e}$) and
also when it is included ($\sigma B_{t \to b \nu e+X}$).
The top quark mass is $m_t=178$ GeV and cross sections are 
evaluated using the CTEQ6M PDF set with all scales equal to
$50$~GeV. The errors represent Monte Carlo statistics only.
Note that the values of $\Gamma_t$ at LO and NLO are $1.651$~GeV and
$1.480$~GeV respectively and the branching ratio of the $W$ into
leptons is ${\rm Br}(W \to e \nu) = 0.1105$.
\label{tab:decay}}
\begin{center}
\begin{ruledtabular}
\begin{tabular}{lllll}
Collider, $\sqrt{s}$ &
$\sigma_0 B_{t \to b \nu e}$ [fb] &
$\sigma B_{t \to b \nu e}$ [fb] &
$\sigma B_{t \to b \nu e+X}$ [fb] \\
\hline
$p\bar p$, 1.96 TeV    & 0.8564 $\pm$ 0.0006  & 0.7887 $\pm$ 0.0005 & 0.7806 $\pm$ 0.0005 \\
$pp$,        14 TeV    & 356.9  $\pm$ 0.2     & 391.7  $\pm$ 0.3    & 395.7 $\pm$ 0.3 \\
\end{tabular}
\end{ruledtabular}
\end{center}
\end{table*}
We have used the CTEQ6M PDF set for all the cross sections in this table, so that
the effect of including radiation in the decay can be understood more easily.
When including radiation in the decay, the total cross section should change by
an amount that is formally of higher order in $\alpha_S$ and is given 
by~\cite{Campbell:2004ch},
\beq
\sigma B_{t \to b \nu e+X} - \sigma B_{t \to b \nu e} =
\left( \frac{\Gamma_t^{LO}}{\Gamma_t^{NLO}}-1 \right)
\left( \sigma B_{t \to b \nu e} - \sigma_0 B_{t \to b \nu e}\right).
\eeq
The results shown in the table agree with this expectation. Since, in our approach,
the effect of the NLO corrections in the production is fairly small,
the numerical difference is only $1$\% at both colliders.

\begin{figure}
\begin{center}
\includegraphics[angle=90,width=0.48\columnwidth]{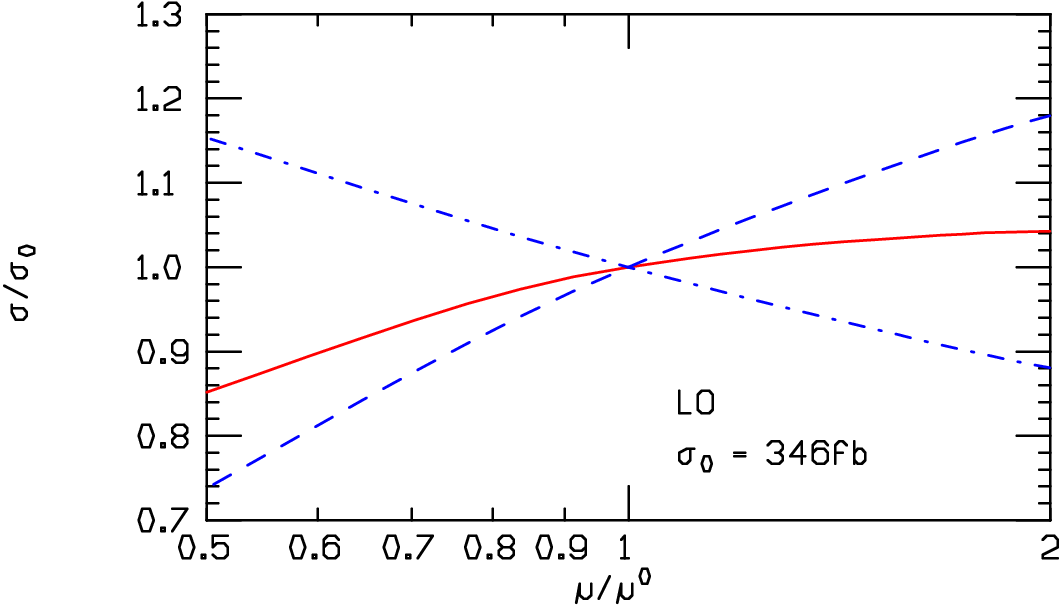}
\hspace*{0.3cm}
\includegraphics[angle=90,width=0.48\columnwidth]{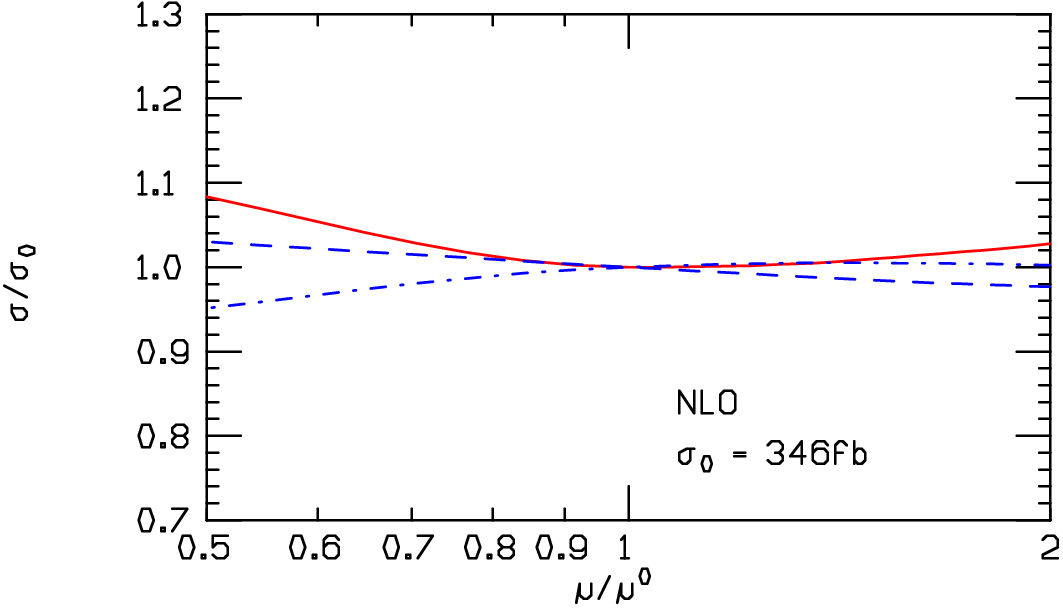}
\caption{Scale dependence of the cross sections for
$W^-(\to e^- {\bar \nu})t(\to \nu e^+ b)$ production at the LHC,
at LO (left) and NLO (right). The NLO calculation includes the effect
of radiation in the decay of the top quark. 
The scale $\mu$ is expressed as the ratio with the central
scale $\mu_0=\ptbv=30$~GeV and the cross sections are scaled to
the central result at $\mu=\mu_0$. 
The solid lines represent the variation of both renormalization
and factorization scales together ($\mu_R=\mu_F=\mu$), the dashed ones the result when only
$\mu_F$ is varied ($\mu_R=\mu_0$) and the dot-dashed curves represent the dependence on the
renormalization scale alone ($\mu_F=\mu_0$).
\label{fig:mudepcteq6dk}}
\end{center}
\end{figure}

We conclude this section with a more detailed presentation of the scale
dependence of our calculation when using the updated parameters and PDF set.
Anticipating the study of the following section, we also choose a lower
value for the ${\bar b}$ jet veto in  the next-to-leading order calculation,
$\ptbv=30$~GeV. This results in the scale dependence shown in
Fig.~\ref{fig:mudepcteq6dk}, where the cross sections at each choice of scale
are expressed as a ratio with the central result at $\mu_0=\ptbv$ and
we vary the scales by a factor of two about $\mu_0$. We also
show the curves obtained when varying the renormalization and factorization
scales separately.  One sees that the relatively small scale dependence at
lowest order is the result of a large cancellation between the dependence on
the factorization and renormalization scales individually. In contrast, the
dependence on these scales at next-to-leading order, either on their own or
when varied together, is small -- less than $10$\% over this range. We also
note that, for this choice of parameters and veto, the
next-to-leading order corrections do not alter the tree level cross
section, which remains at $346$~fb.

\section{The $Wt$ background to $H \to WW^\star$}
\label{sec:pheno}

As an example of the utility of our calculation, in this section we
consider the effect of NLO corrections in the context of a search for
the Higgs boson at the LHC. In the intermediate mass range, $155 < m_H
< 180$~GeV, one of the search strategies involves Higgs production via
gluon fusion, with the subsequent decay of the Higgs boson into
off-shell $W$ pairs which then decay
leptonically~\cite{Dittmar:1996ss},
\beq
\bothdk{g+g \to H}{W^-+}{W^+}{\nu+e^+}{e^-+{\bar \nu}}
\eeq
The largest background in this
channel is from the continuum production of $W$ pairs, both from
diboson production via quark-antiquark scattering and from loop-induced gluon-gluon
fusion~\cite{Binoth:2005ua,Duhrssen:2005bz}. A further significant
source of background events comes from processes producing top-quarks
that decay leptonically. Since the presence of neutrinos in the signal
prevents a full reconstruction of the Higgs mass peak, an accurate
prediction of all the backgrounds is necessary.

In previous studies~\cite{AtlasTDR,Davatz:2004zg} two top backgrounds
have been considered using Pythia~\cite{Sjostrand:2000wi}.
These are resonant $t{\bar t}$
production and $Wt$ production, the process that we consider here. As
we have discussed previously, these two processes become entangled at
NLO. We will separate them according to the procedure that we
outlined in Section~\ref{sec:separation}. Therefore, in addition to our
NLO calculation of the $Wt$ process, we will also consider the contribution
from resonant $t{\bar t}$ production, with the appropriate top quark
decays.

Since the signal process contains no jets at lowest order, it is 
efficient to impose a veto on all jet activity to reduce the size
of these backgrounds.
Our application of a cut on the transverse momentum of the ${\bar b}$
jet in the $Wt$ process fits naturally into this procedure.
We simply extend our veto
to disallow all contributions with any jet observed above the veto
threshold. We note that there is a slight mismatch due to the fact
that our theoretically-motivated veto applies at all rapidity values
whilst the experimental approach only vetoes a jet up to a few units of
rapidity. However, we do not expect this to greatly affect our results.

For our parton-level study we adopt a minimal set of cuts and
examine the effect of the NLO corrections on a selection of observables
that are typically used in more detailed experimental
studies~\cite{Dittmar:1996ss,AtlasTDR,Davatz:2004zg,Duhrssen:2005bz}.
Our basic cuts represent the finite acceptance limits of the detectors
at the LHC,
\beq
p_T({\rm lepton}) > 20~{\rm GeV} , \qquad
|\eta({\rm lepton})| < 2.5,
\label{eq:cutbegin}
\eeq
applied to both of the leptons produced in the $W$ decays, with the missing
transverse momentum also constrained by,
\beq
p_T({\rm missing}) > 30~{\rm GeV}.
\eeq
The final cut that we apply is the jet veto, after potential jets have
been clustered according to the $k_T$ algorithm with a jet separation
parameter $\Delta R=1.0$. Events are not included if any jet is
observed with,
\beq
p_T({\rm jet}) > 30~{\rm GeV} , \qquad
|\eta({\rm jet})| < 3.
\label{eq:cutend}
\eeq 
We note that, in addition to excluding additional radiation at 
next-to-leading order, this veto also applies to the $b$ jet that is
produced in the top quark decay. This results in a substantial
decrease in cross section compared to the totally inclusive case.

\begin{figure}
\begin{center}
\includegraphics[angle=90,width=0.8\columnwidth]{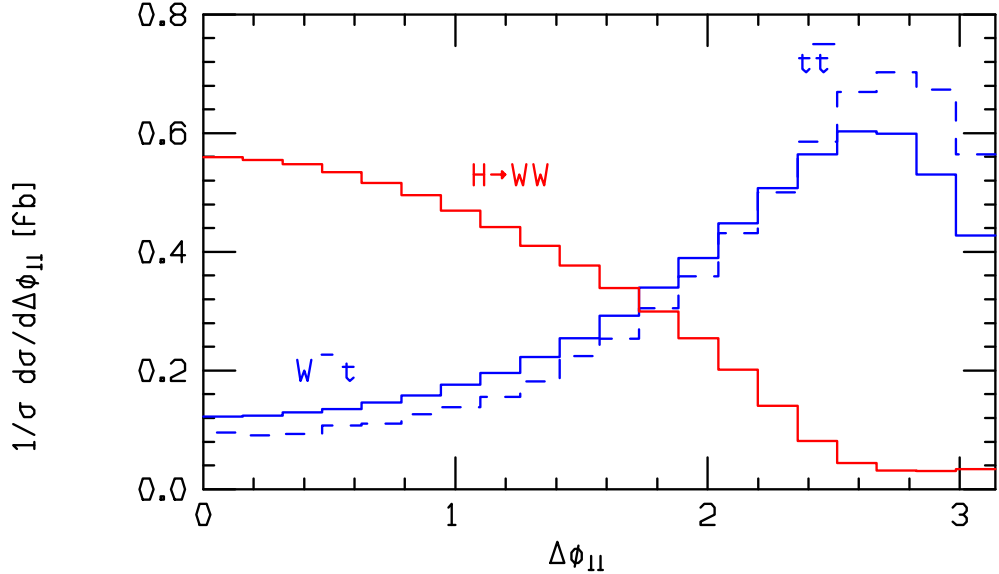}
\caption{
The distribution of the opening angle between the leptons in the
transverse plane, for the signal and the two background processes
considered here. All curves are lowest order predictions and are
normalized to unity. The signal calculation uses a Higgs mass of
$155$~GeV and the backgrounds are $W^-t$ (solid) and $t{\bar t}$
(dashed).
\label{fig:dphicompare}}
\end{center}
\end{figure}

To exploit the spin correlation between the leptons in the signal
events~\cite{Nelson:1986ki}, one can make quite stringent cuts 
on the opening angle between the leptons in the transverse plane,
$\Delta \phi_{\ell \ell}$. This is illustrated in
Fig.~\ref{fig:dphicompare}, where we show the shapes of the lowest
order predictions for the Higgs signal and the two top backgrounds,
$Wt$ and $t{\bar t}$. Signal events predominantly contain leptons
with a small opening angle between them, whereas both backgrounds tend
to produce leptons that are almost back-to-back in the transverse plane.

\begin{figure}
\begin{center}
\includegraphics[angle=90,width=0.8\columnwidth]{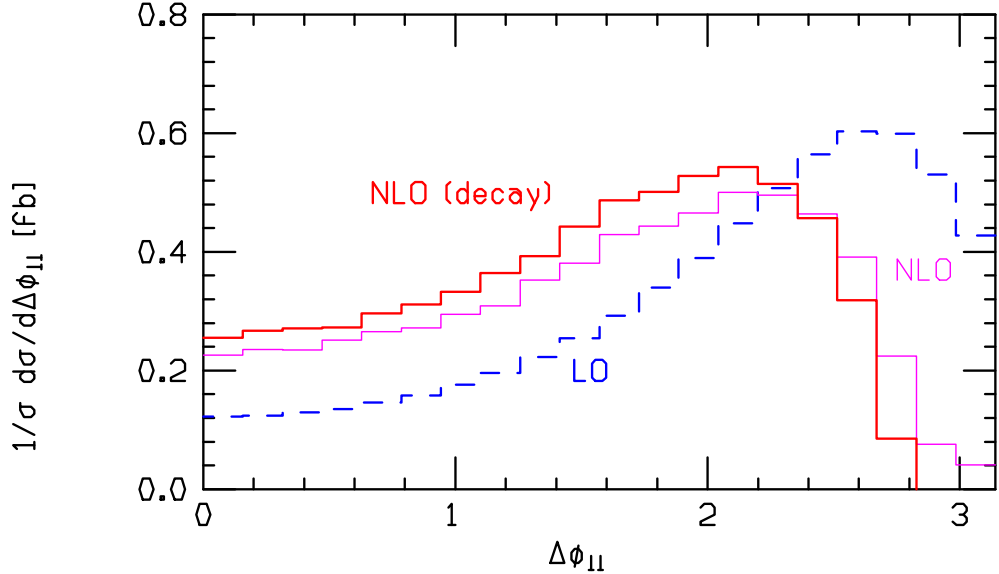}
\caption{
The distribution of the opening angle between the leptons in
the transverse plane for $W^-t$ events at the LHC, calculated at LO (dashed)
and NLO (solid) for the LHC. The NLO curves are labelled according to
whether or not they include the effect of radiation in the decay.
All the rates are normalized to unity.
\label{fig:dphi}}
\end{center}
\end{figure}

We now examine the extent to which this is changed in our NLO
calculation of the $Wt$ background, with our results shown in
Fig.~\ref{fig:dphi}. 
The effect of the NLO corrections is to change the shape of the
distribution considerably when these cuts are applied. The peak at
large $\Delta \phi_{\ell \ell}$ is shifted to a smaller value and
becomes much less pronounced. This could have quite a large impact on a
strategy in which this background is measured using events at large
$\Delta \phi_{\ell \ell}$ and then extrapolated via the theoretical
shape to the Higgs signal region. One also sees that the shape is
changed again when including the effects of radiation in the top quark
decay, with the peak being sharpened once more, although the effect is
fairly minor.

\begin{figure}
\begin{center}
\includegraphics[angle=90,width=0.8\columnwidth]{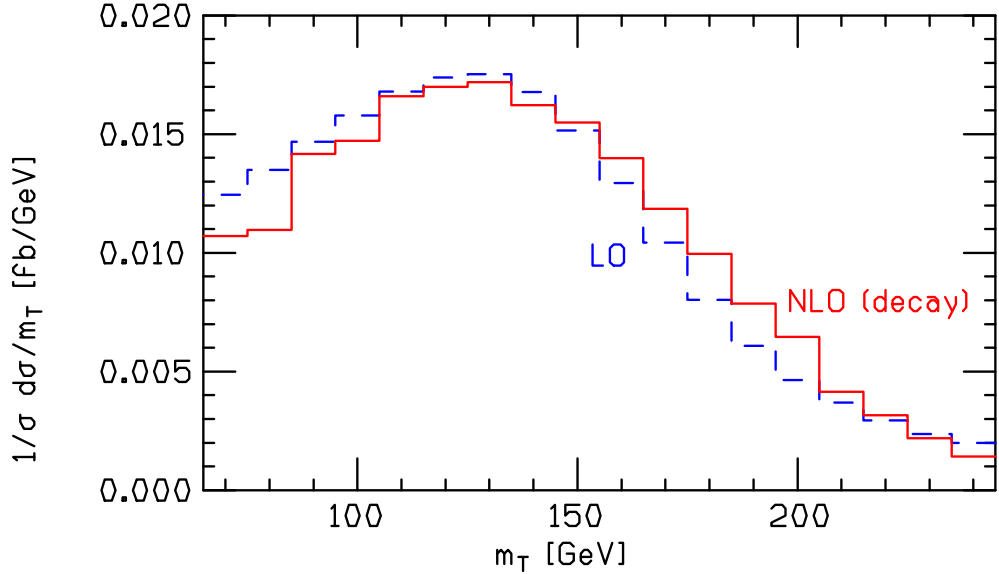}
\caption{
The distribution of the transverse mass for $W^-t$ events at the LHC,
calculated at LO (dashed) and NLO (solid) for the LHC. The NLO result includes
the effect of radiation in the decay. The rates are normalized to unity.
\label{fig:mtrans}}
\end{center}
\end{figure}

One can also imagine constructing the transverse mass of the putative
Higgs boson ($m_T$) from the transverse momenta of the dilepton system
and the missing $p_T$,
\beq
m_T=\sqrt{2 p_T^{\ell \ell} p_T^{\rm miss} (1-\cos(\Delta\phi))},
\eeq
where $\Delta \phi$ is the angle between the two vectors in the
transverse plane. Cutting in a suitable mass window can further help
to reduce the backgrounds for only a small loss in signal. The impact of our
next-to-leading order calculation on this distribution is illustrated
in Fig.~\ref{fig:mtrans}.
In this figure we have only shown the result when including radiation
in the decay but note that this distribution changes little when it is
excluded. One can see that the shape of this distribution is relatively
unchanged at NLO, although more events are produced at high values of
$M_T$, beyond the peak of the distribution, than at LO. 

Finally, to give some idea of the effect of the NLO corrections on
the number of events that should be observed in this channel, in
Table~\ref{tab:cutxsecs} we show the cross sections that we find
for the $Wt$ and $t{\bar t}$ processes. The $W$'s decay into electrons
only and for the $Wt$ process, both $W^-t$ and $W^+{\bar t}$ are
included. Results are shown at LO, NLO and at NLO when including
radiation in the top quark decay. For the $t{\bar t}$ process,
the lowest order diagrams of Fig.~\ref{fig:wtbdiags} (b)
are calculated. As an approximation to the NLO result, a
$K$-factor is applied from the NLO calculation involving no top quark
decay. In order to match the study more closely, we have used
a $K$-factor obtained when applying the jet veto of Eq.~(\ref{eq:cutend}).
Using the common scales $\mu_R=\mu_F=m_t$, we find that this factor is
${\cal K}=0.7$.
\begin{table*}[tb]
\caption{\small Comparison of LO and NLO cross sections for top quark
backgrounds in an intermediate mass Higgs search at the LHC. Results are
shown in femtobarns for ($W^-t+W^+{\bar t}$) and $t{\bar t}$ production,
with leptonic decays of both the $W$ and the top quark.
Three sets of cuts are considered, which are described in
detail in the text. The NLO $Wt$ calculation is performed both
without including QCD effects in the decay ($\sigma B_{t \to b \nu e}$) and
also when it is included ($\sigma B_{t \to b \nu e+X}$).
\label{tab:cutxsecs}}
\begin{center}
\begin{ruledtabular}
\begin{tabular}{lccc}
                             & basic       & $m_H=155$~GeV    & $m_H=180$~GeV \\
\hline
\multicolumn{4}{c}{$Wt$ process}                                              \\
$\sigma_0 B_{t \to b \nu e}$ [fb] & 40.08       & 0.80             & 0.80          \\
$\sigma B_{t \to b \nu e}$ [fb]   & 13.14       & 0.38             & 0.43          \\
$\sigma B_{t \to b \nu e+X}$ [fb] & 13.85       & 0.46             & 0.51          \\
\hline
\multicolumn{4}{c}{$t{\bar t}$ process}                                       \\
$\sigma_{LO}\times{\cal K}$ [fb]  & 30.52       & 0.42             & 0.43          
\end{tabular}
\end{ruledtabular}
\end{center}
\end{table*}

The cross sections shown in Table~\ref{tab:cutxsecs} are evaluated using
three different sets of cuts. The first column uses just the 
basic set of cuts (Eqs.~(\ref{eq:cutbegin})--(\ref{eq:cutend})), then the
other two columns represent extensions of these cuts that might be used
in the search for a Higgs boson of given mass. Both the further sets also impose,
\beqn
\Delta \phi_{\ell \ell} & < & \pi/4 , \nonumber \\
m_{\ell \ell} & < & 35~{\rm GeV},
\eeqn
to select the Higgs signal region. In addition we have used
a cut on the reconstructed transverse mass around the Jacobian peak of
the putative Higgs mass. In the first case,
the cut is constructed for a Higgs mass of $155$~GeV, by
constraining $125 < m_T < 155$~GeV. The second set requires that
$140 < m_T < 180$~GeV and is aimed at a search for a $180$~GeV Higgs.

One sees that the effect of the NLO corrections in this region of phase
space is significantly different from the inclusive case. The cross
section is decreased substantially, with a $K$-factor (for the
calculation including radiation in the decay) of approximately $0.6$
when applying the Higgs search cuts. We also see that the contributions
from the two top quark processes are comparable when calculated in this
way.

\section{Conclusions}

At the LHC, the top quark will be readily produced in association with
a $W$ boson. In this paper we have performed a next-to-leading order
calculation of this process, including both the subsequent leptonic
decays $W^- \to e^- {\bar \nu}$ and $t \to \nu e^+ b$, as well as
the emission of real radiation in the top decay.

For total inclusive cross sections, where the top quark and $W^-$ boson
do not decay, comparison with previous calculations is possible and we
find results which are broadly similar. However, due to the presence of
doubly-resonant $t{\bar t}$ diagrams at NLO -- which are handled differently
in our calculation -- we do not find exact agreement. Our method
maintains the consistency of the $b$-PDF approach and requires the
simultaneous use of a relatively low factorization scale $\mu_F \sim
(m_t+m_W)/4$ and a veto on ${\bar b}$ quarks with a transverse momentum
larger than this value of $\mu_F$.

With this approach, we find that the NLO corrections to the inclusive rate
at the Tevatron can be large, but the process remains phenomenologically
irrelevant there. At the LHC the corrections are smaller and can either
increase or decrease the cross section by $10$\%, depending on the
${\bar b}$ quark veto that is chosen. We also performed an
analysis of this process as a background to Higgs production at the LHC,
where quite a severe veto on all jet activity is applied.
We find that the shape of the $\Delta \phi_{\ell \ell}$ distribution
-- crucial to the estimation of this background -- is significantly changed.
However the cross section for this process
can be reduced by as much as $40$\% in the Higgs signal region.

Our predictions are based on a number of approximations. Firstly, the
mass of the $b$ quark is set to zero throughout and the top quark is
kept on its mass shell. The inclusion of the $b$-mass in the top quark
decay and implementation of a Breit-Wigner distribution for the top
quark can be studied in the lowest order calculation. Neither of these
significantly alters the total cross section or the shapes of the
distributions which we have examined. Effects due to the interference
between radiation in the production and decay stages are estimated to
be similarly small, of order $\alpha_s \Gamma_t/m_t$. Lastly, our
predictions are performed at parton-level only and lack any modelling
of hadronization and showering. This is particularly relevant to our
phenomenological analysis, where the application of a jet veto at
relatively low transverse momenta (compared to the top quark mass)
increases the sensivity to soft-gluon effects. Nevertheless, we hope
that our results can provide a starting point for further investigation
of the $Wt$ process at next-to-leading order accuracy.

\begin{acknowledgments}
We thank K.~Ellis for collaboration in the early stages of this work and
we would like to acknowledge many helpful discussions with F.~Maltoni
and S.~Willenbrock.
\end{acknowledgments}

\appendix

\section{Integration of dipoles}
\label{appendix:real}

We have slightly extended the approach outlined in Appendix A of
Ref.~\cite{Campbell:2004ch} in order to handle the presence of
gluons in the initial state. We refer the reader to
Ref.~\cite{Campbell:2004ch} and the original papers
CS~\cite{Catani:1997vz} and CDST~\cite{Catani:2002hc} for further
explanation of the method and notation.

\subsection{Initial-state emitter with initial state spectator}
As explained in Ref.~\cite{Campbell:2004ch},
we have generalized the dipole
phase space (CS, Eq.~(5.151)) by introducing a constraint
enforced by the factor $\Theta(\alpha-{\tilde v}_i)$.
The variable ${\tilde v_i}$ is the rescaled value of the propagator defined
by,
\beq   
{\tilde v_i}= \frac{p_a p_i}{p_a p_b}
\eeq   
where $p_a$ is the initial state emitter, $p_i$ is the emitted parton and 
$p_b$ is the other initial state parton which is the spectator. 
Further details are given in section 5.5 of CS.
In this appendix we extend the previous treatment by considering the
remaining $q,g$ and $g,g$ cases. The dipole integrands which we
subtract are obtained by modifying CS, Eq.~(5.154):
\beq
\langle\,
{\bom V}^{q_aq_i,b}(x_{i,ab}) \,\rangle
= 8\pi \mu^{2\ep} \as\; C_F\;
\left[ \frac{1 + (1 - x_{i,ab})^2}{x_{i,ab}} -\eta\,\ep\,x_{i,ab}\right]
\;\delta_{ss'} \;,
\eeq
\beq
\langle\,
{\bom V}^{g_ag_i,b}(x_{i,ab}) \,\rangle
= 8\pi \mu^{2\ep} \as\; 2\,C_A\;\left[ \frac{x_{i,ab}}{1-x_{i,ab}}
+ \frac{1-x_{i,ab}}{x_{i,ab}} + x_{i,ab}\,(1- x_{i,ab}) \right]
\; .
\eeq

We find that the result for the $q,g$ case is given by,
\beqn
&&{\tilde \cV}^{q,g}(x;\ep,\alpha)
=C_F \Bigg\{ \frac{\Big( 1+(1-x)^2 \Big)}{x} \nn \\
&\times & \Bigg[2 \ln(1-x)-\frac{1}{\ep}
    +\Theta(1-x-\alpha) \ln\bigg(\frac{\alpha}{1-x}\bigg)\Bigg]
  + \eta\,x \Bigg\} +\Oe{}\: ,
\eeqn
and for the $g,g$ case it is,
\beqn
&&{\tilde \cV}^{g,g}(x;\ep,\alpha)
=C_A \Bigg\{\Bigg(\frac{1}{\ep^2}-\frac{\pi^2}{6}\Bigg) 
  \delta(1-x) \nn \\
&+& 2\Big( x\,(1-x)+\frac{1-x}{x} -1 \Big) \Big(2 \ln(1-x)-\frac{1}{\ep}\Big)
\nn \\
&+&\Theta(1-x-\alpha) \frac{2(x\,(1-x)-1)^2}{x(1-x)} 
 \ln\Big(\frac{\alpha}{1-x}\Big) 
\nn \\
      &-& \frac{2}{\ep} \frac{1}{\big[1-x\big]_{+}}
  +4 \Bigg[\frac{\ln(1-x)}{1-x}\Bigg]_{+} \Bigg\} +\Oe{}\: .
\eeqn

In the limit $\alpha=1$ these functions correspond to
those given in CS, Eq.~(5.155) and also agree with the results of
Nagy~\cite{Nagy:1998bb,Nagy:2003tz}.

\subsection{Initial-state emitter with final-state spectator}

In this category, we complete the treatment of Ref.~\cite{Campbell:2004ch}
by considering the remaining three cases:
\beqn
&& \mbox{a) Initial } g  \to q+\bar{q} , \nn \\
&& \mbox{b) Initial } q  \to g+q , \nn \\
&& \mbox{c) Initial } g  \to g+g ,  
\eeqn
where the first parton on the right-hand side is an initial-state
emitter and the last one is a massive final-state
spectator. The phase space is the generalization of CDST, Eq.~(5.79) with an
extra factor of $\Theta(\alpha-z_i)$.

For case (a) the dipole integrand is given by a generalization of 
CDST, Eq.~(5.82). The result is written in the terms of the variable
$z_+$ defined by,
\beq
z_+=\frac{1-x}{1-x+\mu_Q^2},
\eeq
and we find,
\def\mut{\tilde \mu}
\beqn
\label{if1}
&&I^{gq}(x;\ep,\alpha)
=T_R \Bigg\{ \Big( x^2 + (1-x)^2 \Big) \nn \\
&\times & \Bigg[ \ln\Big(\frac{(1-x)^2}{1-x+x\mut^2}\Big) -\frac{1}{\ep}
 - \Theta(z_+ -\alpha) \ln\bigg(\frac{z_+}{\alpha}\bigg)\Bigg]
 +2\,\eta\,x(1-x)  \Bigg\} +\Oe{}\: . 
\eeqn
In this expression we have also introduced the variable,
\beq
{\tilde \mu^2} = \frac{\mu^2}{x} = \frac{m^2}{2 {\tilde p_{ai}} {\tilde p_j}} ,
\eeq
which only depends on ${\tilde p_{ai}}$ and  ${\tilde p_j}$ (defined in
CDST Eq.~(5.73)), the momenta held fixed
when the $x$ integration is performed.

For case (b) the integrand is obtained by generalizing Eq.~(5.84) of
CDST. Performing the integrals yields the result,
\beqn
\label{if2}
&&I^{qg}_Q(x;\ep,\alpha)
=C_F \Bigg\{ \frac{\Big( 1+(1-x)^2 \Big)}{x} \nn \\
&\times & \Bigg[ \ln\Big(\frac{(1-x)^2}{1-x+x\mut^2}\Big) -\frac{1}{\ep}
 -\Theta(z_+ -\alpha) \ln\bigg(\frac{z_+}{\alpha}\bigg)\Bigg]
   + 2\,\mut^2 \ln \bigg(\frac{x\mut^2}{1-x+x\mut^2}\bigg) \nn \\
&-& \Theta(z_+ -\alpha) 2 \mut^2\,\ln\bigg(\frac{1-z_+}{1-\alpha}\bigg)\Bigg]
   + \eta\,x \Bigg\} +\Oe{}\: .
\eeqn

Lastly, the dipole integrand for case (c) is an extension of 
CDST, Eq.~(5.86). The result in this case is,
\beqn
\label{if3}
&&I^{gg}_Q(x;\ep,\alpha)
= \,C_A \Bigg\{
  \delta(1-x) \Bigg[ \frac{1}{\ep^2}+\frac{1}{\ep}\ln(1+\mut^2)+\frac{\pi^2}{6} \nn \\
&+& 2 \Li_{2}(-\mut^2) +2 \ln(\mut^2) \ln(1+\mut^2)- \frac{1}{2} \ln^2(1+\mut^2) \Bigg] \nn \\
&-&\frac{1}{\ep}\,\frac{2}{[1-x]_{+}} -2\;\frac{\ln(1+{\tilde \mu}^2)}{[1-x]_{+}} 
 +4\;\Big[\frac{\ln(1-x)}{(1-x)}\Big]_{+} \nn \\
 &+& 2\,\Big( x(1 - x) + \frac{1 - x}{x} - 1 \Big) \Bigg[ - \frac{1}{\ep}
 + \ln\Big(\frac{(1-x)^2}{1-x+x\mut^2}\Big) 
 - \Theta(z_+-\alpha)\,\ln\Big(\frac{z_+}{\alpha}\Big) \Bigg] \nn \\
 &-& \Theta(z_+-\alpha) \Bigg[ 
  2 \mut^2\,\ln\Big(\frac{1-z_+}{1-\alpha}\Big)
 - \frac{2}{(1-x)} \ln\Big(\frac{\alpha(1-x+z_+)}{z_+(1-x+\alpha)}\Big) \Bigg] \nn \\
&+& 2\mut^2\,\ln\Big(\frac{x\mut^2}{1-x+x\mut^2} \Big) 
 - \frac{2}{(1-x)} \ln\Big(\frac{2-x+x\mut^2}{1+\mut^2} \Big)
\Bigg\} .
\eeqn

\section{The virtual $++$ amplitude}
\label{appendix:virt}

We remind the reader that we calculate the virtual amplitudes for
the production of a $W$ that decays leptonically in association with
an on-shell top quark,
\beq
b+g \longrightarrow W (\to e+n) + t,
\eeq
where $n$ represents the neutrino, $t^2=m_t^2$ and
$b^2=g^2=e^2=n^2=0$.

We first introduce the following short-hand notation,
\beqn
&&\tau_{xy}=2 \, x \cdot y ~,~
t_{xy}=(x+y)^2 ~,~
\qsq=q^2=2\,e \cdot n \nn \\
&&q_{xy}=\qsq-t_{xy} ~,~
\qsqhat=\qsq-\mt^2 \nn \\
&&\delta_{tb}=\la b\,t_0 \ra \, [t_0\,b] = 2\,b \cdot t_0
= 2\,b \cdot t - m_t^2\,\frac{b \cdot g}{t \cdot g},
\label{eq:topdecomp}
\eeqn
such that the massive top quark momentum $t$ does not appear
directly in the following formulae but is instead replaced by
the massless vector $t_0$, which is suitable for our spinor
approach.

Next we define a set of functions that are useful for decomposing the
form of the virtual amplitudes. In these formulae, the
scalar $n-$point functions are written using the following definition,
\beq
I_0^n(\{p_{1...n-1}\};\{m_{1...n}^2\})=
\frac{\mu^{2\ep}}{r_\Gamma}\int\frac{d^{4-2\ep}l}{(2\pi)^{4-2\ep}}
\frac{1}{(l^2-m_1^2)((l-p_1)^2-m_2^2)~...~((l-p_{n-1})^2-m_n^2)},
\eeq
where,
\beq
r_\Gamma=\left(\frac{4\,\pi\,\mu^2}{m_t^2}\right)^{\ep}
 \frac{i}{16\,\pi^2\,\Gamma(1-\ep)}.
\eeq
We subsequently identify $I_0^2( \ldots )\equiv B_0( \ldots )$,
$I_0^3( \ldots )\equiv C_0( \ldots )$ and $I_0^4( \ldots )\equiv D_0( \ldots )$.
The simplest functions contain only bubble integrals,
\beqn
l_Q&=& 1/\ep+2-B_0(q;0,\msq) = \qsqhat \, \log( -\qsqhat/ \msq ) / \qsq \nn \\ 
\ltg &=& 1/\ep+2-B_0(t+g;0,\msq) = \tautg  \, \log( -\tautg / \msq ) / \ttg  \nn \\ 
\ltb &=& 1/\ep+2-B_0(t+b;0,\msq) = \tautb  \, \log( -\tautb / \msq ) / \ttb  \nn \\ 
l^1_{Q} &=& \qsq \, (\ltg-l_Q) / \qtg  \nn \\
l^1_{tb} &=& \ttb \, (\ltb-l_Q) / \qtb  \nn \\
\l^2_{tg} &=&
      (\ltg-l_Q)\,\qsq^2/\qtg^2
    + l_Q\,\qsq^2/\qtg/\tautg
    - l_Q\,\qsq/\qtg
    + \qsq/\qtg
\eeqn
Three triangle functions appear in our results, which we choose to
keep as independent functions,
\beqn
C_0^A&=&C_0(t+b,g;0,\msq,\msq)\nn \\
C_0^B&=&C_0(t,g;0,\msq,\msq) \nn \\
C_0^C&=&C_0(g+b,t;0,0,\msq)
\eeqn
Lastly, we form functions which are combinations of the triangle
functions and the basic box integrals that enter our calculation,
\beqn
L^6_1&=&\tautb\,\taugb\,(D_0(g,b,t;0,0,0,\msq)
  -C_0(g,b;0,0,0)/\tautb  \nn \\
&&  +C_0(g,t+b;0,0,\msq)\,(\taugb+\tautg)/\taugb/\tautb
  -C_0(b,t;0,0,\msq)/\taugb  \nn \\
&&  +C_0(g+b,t;0,0,\msq)\,(\tautb\,(\tautg+\tautb)-2\,m_t^2\,\taugb)/\taugb/\tautb^2) ,
\eeqn
\beqn
L^6_2&=&\tautb\,\tautg\,(D_0(b,t,g;0,0,\msq,\msq)
   -C_0(b,t;0,0,\msq)/\tautg  \nn \\
&&   +C_0(b,t+g;0,0,\msq)\,(\taugb+\tautb)/\tautg/\tautb
   +C_0(t,g;0,\msq,\msq)\,(2\,m_t^2\,\taugb-\tautg\,\tautb)/\tautg/\tautb^2  \nn \nn \\
&&   +C_0(t+b,g;0,\msq,\msq)\,(\tautg+\taugb)
                          \,(\tautg\,\tautb-2\,m_t^2\,\taugb)/\tautg^2/\tautb^2) ,
\eeqn
\beqn
L^6_3&=&\tautg\,\taugb\,(D_0(b,g,t;0,0,0,\msq)
  -C_0(b,g;0,0,0)/\tautg  \nn \\
&&  +C_0(b,t+g;0,0,\msq)\,(\taugb+\tautb)/\taugb/\tautg
  -C_0(g,t;0,0,\msq)/\taugb  \nn \\
&&  +C_0(g+b,t;0,0,\msq)\,(\tautg\,(\tautg+\tautb)-2\,m_t^2\,\taugb)/\taugb/\tautg^2) .
\eeqn
All the basic scalar integrals, $B_0$, $C_0$ and $D_0$ are well
known~\cite{'tHooft:1978xw,Beenakker:2002nc}.

To renormalize the virtual amplitudes
we have used the modification of the $\overline{MS}$ scheme in which the top
quark is decoupled. The top self-energy is renormalized on-shell
so that we have evaluated the self energy in the top-right diagram of
Fig.~\ref{fig:virtp} including the following mass counterterm,
\beq
\delta Z_{mass,c.t.}=-\frac{\alpha_s}{4\,\pi}
\left( \frac{4\,\pi\,\mu^2}{m_t^2}\right)^\ep
\frac{C_F\,m_t}{\Gamma(1-\ep)}\,
\left(\frac{3}{\ep}+5-\eta\right)\,\bar{\psi}_t\psi_t,
\eeq
and we renormalize the top wave function by adding,
\beq
\frac{\delta Z_{wf}}{2}=-\frac{\alpha_s}{4\,\pi}
\left( \frac{4\,\pi\,\mu^2}{m_t^2}\right)^\ep
\frac{C_F}{2\,\Gamma(1-\ep)}\,
\left(\frac{3}{\ep}+5-\eta\right)\,A_{Tree~Level}.
\eeq
In the above formulae $\eta$ is a parameter that specifies the
regularization scheme adopted throughout the calculation.
$\eta=0$ corresponds to the 4-dimensional helicity scheme, which 
we use here, whilst $\eta=1$ is appropriate in the 't Hooft Veltman
scheme. Subtracting the top loop contribution to the gluon self energy
at zero momentum transfer enables the coupling constant to evolve due to
the presence of 5 light flavours only. Finally, the partial cancellation of
the coupling constant renormalization and this top contribution to the
gluon self energy give us the last contribution,
\beq
\delta Z_{g}'=-\frac{\alpha_s}{4\,\pi}(4\,\pi)^\ep
\frac{(11N/6-2T_R\,n_f/3)}{\ep\,\Gamma(1-\ep)}\,
A_{Tree~Level}.
\eeq

We are now in a position to write down the amplitudes for the specific
helicity choice `++'. The first `+' signifies the helicity of the gluon
and the second `+' the spin of the top quark in a basis determined by
our decomposition in Eq.~(\ref{eq:topdecomp}). The tree-level amplitude
is,
\beq
A_{Tree~Level}^{++}=
 \frac{\sqrt{2}\,g_s\,g_W^2\,T^a}{\,W^-_{prop}}\,A_{0}^{++},
\eeq
where,
\beq
A_{0}^{++}=-\,\mt\,\frac{\la g\,e \ra}{\la g\,b \ra\,\la g\,t_0 \ra^2}\,
 (\la b\,t_0 \ra\,[b\,n]+\la g\,t_0 \ra\,[g\,n]),
\eeq
and,
\beq
W^-_{prop}=q^2-M_W^2+i M_W \Gamma_W.
\eeq
The corresponding virtual amplitude is decomposed into a piece containing
poles in $\epsilon$ which is proportional to the lowest order result, plus a finite
remainder,
\beqn
A_{Virtual}^{++}&=&\frac{\alpha_s}{4\,\pi}\frac{\sqrt{2}\,g_s\,g_W^2\,T^a}{\Gamma(1-\ep)\,W^-_{prop}}
\left(\frac{4\,\pi\,\mu^2}{m_t^2}\right)^{\ep}
\left[(F+P_{wf}) \cdot A_{0}^{++}+A_{1}^{++}\right].  
\eeqn
The pole pieces are given by,
\beqn
F&=&
   -N\,(3/2/\ep^2-1/\ep\,\log(-\tautg/\msq)-1/\ep\,\log(-\taugb/\msq)  \nn \\
&&  +1/2/\ep+\log(-\tautg/\msq)^2+1/2\,\log(-\taugb/\msq)^2) \nn \\
&&   +1/2/N\,(1/\ep^2-2/\ep\,\log(-\tautb/\msq)+1/\ep+2\,\log(-\tautb/\msq)^2)  \nn \\
&&   -N\,(\Li_{2}(\ttg/\msq)+\,\pi/12)+(\Li_{2}(\ttb/\msq)+\,\pi^2/12)/N,  
\eeqn
while the term which is the sum of coupling constant and wave-function renormalization is,
\beq
P_{wf}=(2\,T_R\,n_f/3-11\,N/6)\,(1/\ep-\log(\mu^2/\msq))
-3C_F/2\,(1/\ep+5/3). 
\eeq
The remainder is then written in terms of our functions as,
\beqn
&&A_{1}^{++}=
\frac{\mt\,\la g\,e \ra \, [g\,n]}{\la g\,b \ra \, \la g\,t_0 \ra \, \la b\,t_0 \ra \, [b\,t_0]}\,\Big\{
	\frac{1}{\xn}\,\big[
            L^6_1\,\delta_{tb}/2
          + L^6_2 \left( \delta_{tb}/2 - \qtg\,\delta_{tb}/\taugb - \qtg\,\mt^2/2/\tautg \right) \nn \\
&&          - C_0^A\,\taugb\,\delta_{tb}\,\mt^2/\tautg
          - 2\,C_0^A\,\delta_{tb}\,\mt^2
          + 2\,C_0^B\,\delta_{tb}\,\mt^2
          + C_0^C\,\taugb\,\delta_{tb}\,\mt^2/\tautb
          + l^1_{tb}\,\taugb\,\ttb/\tautg 
          + l_Q\,\delta_{tb} \nn \\
&&        - \ltg\,\ttg\,\delta_{tb}/\tautg
          - \ltb\,\ttb\,\tautb/\tautg
          + \ltb\,\taugb\,\delta_{tb}/\tautb
          + l_Q\,\tautb\,\ttb/\tautg
 \big]
	+C_F\,\big[
            \ltg\,\qtb\,\ttg/\tautg
          - \ltg\,\qsq 
          - \delta_{tb} \nn \\
&&          + \ltg\,\delta_{tb}
          + 2\,\ltg\,\ttg\,\delta_{tb}/\tautg
          + \qtg\,\delta_{tb}/\tautg
          - 2\,\delta_{tb}\,\qsqhat/\tautg
 \big]
	+\xn\,\big[
          - L^6_3\,\delta_{tb}/2
          - C_0^C\,\taugb\,\delta_{tb}\,\mt^2/\tautg
 \big]
 \Big\}  \nn \\
&& + \frac{\mt\,\la g\,e \ra \, [b\,n]}{\la g\,b \ra \, \la g\,t_0 \ra^2 \, [b\,t_0]} \, \Big\{
	\frac{1}{\xn}\,\big[
            L^6_1\,\tautb/2
          + L^6_2 \left( \delta_{tb} - \ttb/2 + \tautb\,\tautg/\taugb \right)          
          + 2\,C_0^A\,\qtb\,\taugb\,\delta_{tb}\,\mt^2/\tautb/\tautg  \nn \\
&&        + C_0^A\,\qtb\,\delta_{tb}\mt^2/\tautb
          - C_0^B\,\qtb\,\delta_{tb}\mt^2/\tautb
          - C_0^B\,\taugb\,\delta_{tb}\mt^2/\tautb
          - 2\,C_0^C\,\taugb\,\delta_{tb}\,\mt^2/\tautg
          - \ltb\,\tautg\,\delta_{tb}/\tautb  \nn \\
&&          - \ltg\,\delta_{tb} \big]
	+C_F\,\big[
          - 4\,C_0^C\,\taugb\,\delta_{tb}\,\mt^2/\tautg
          + 2\,l^1_{Q}\,\taugb\,\qsq/\tautg
          + l^1_{Q}\,\taugb
          - 2\,\ltg\,\qtb
          - 2\,\ltg\,\qtb\,\qsq/\tautg  \nn \\
&&          + \ltg\,\tautg\,\qsq/\ttg
          + l_Q\,\qtg\,\qtb/\tautg
          + l_Q\,\qsq\,\qtb/\tautg
          + l_Q\,\qtb
          + l_Q\,\qsq
          - \qtg\,\tautg/\ttg
          + \taugb
          - \delta_{tb}
 \big]
 \nn \\
&&	+\xn\,\big[
            L^6_3\,\tautb/2
          - L^6_3\,\delta_{tb}
 \big]
 \Big\}
 \nn \\
&& + \frac{\mt\,\la b\,e \ra \, [b\,n]}{\la g\,b \ra \, \la g\,t_0 \ra \, \la b\,t_0 \ra \, [b\,t_0]} \, \Big\{
	\frac{1}{\xn}\,\big[
          - L^6_2\,\qtb\,\mt^2/2/\tautg
          + C_0^A\,\qtb^2\,\delta_{tb}\,\mt^2/\tautb/\tautg
          - C_0^B\,\qtb\,\delta_{tb}\,\mt^2/\tautb  \nn \\
&&          - l^1_{Q}\,\taugb\,\qsq/\tautg
          + \ltb\,\qsq\,\delta_{tb}/\tautb
          - l_Q\,\qsq\,\qtb/\tautg
          + \ltg\,\qtb\,\qsq/\tautg
 \big]
	+C_F\,\big[
            \ltg^2\,\taugb^2/\tautg
          + 2\,l^1_{Q}\,\taugb^2\,\qsq/\tautg^2  \nn \\
&&          - 3\,l^1_{Q}\,\taugb^2/\tautg
          - 2\,l^1_{Q}\,\taugb
          + 4\,\ltg\,\qtb\,\qsq/\tautg
          - 2\,\ltg\,\qtb^2\,\qsq/\tautg^2
          - 2\,\ltg\,\ttg\,\qtb^2/\tautg^2
          + 2\,\ltg\,\qtb^2/\tautg  \nn \\
&&          - \ltg\,\tautg\,\qsq/\ttg
          + l_Q\,\qtg\,\qtb^2\,\mt^2/\tautg^2/\qsqhat
          + l_Q\,\qsq\,\qtg\,\qtb^2/\tautg^2/\qsqhat
          - 2\,l_Q\,\qsq\,\qtb/\tautg
          + l_Q\,\qtb^2\,\mt^2/\tautg/\qsqhat  \nn \\
&&          + l_Q\,\qsq\,\qtb^2/\qsqhat/\tautg
          - \qtb\,\delta_{tb}/\tautg
          - \qtb^2/\tautg
          + \tautg\,\qsq/\ttg
 \big]
 \Big\}.
\eeqn
The expressions for the other amplitudes are similar but slightly
longer, requiring the addition of a further 5 functions to describe them
compactly. We do not reproduce them here, but they are available as Fortran
files from the authors on request.

\end{document}